\documentclass[lettersize,journal]{IEEEtran}
\IEEEoverridecommandlockouts
\usepackage{amsmath,amssymb,amsfonts}
\usepackage{cite}

\usepackage{algorithm}
\usepackage{algpseudocode}
\usepackage{stfloats}
\usepackage{graphicx}
\usepackage{textcomp}
\usepackage{xcolor}
\usepackage{color}
\usepackage{pifont}
\usepackage{amsthm}

\usepackage{hyperref}
\usepackage{mathrsfs}
\usepackage{booktabs}
\usepackage{hyperref}
\usepackage{cases}
\usepackage{comment}
\usepackage{empheq} 
\usepackage{caption}
\usepackage{subcaption}
\usepackage{epstopdf}

\allowdisplaybreaks[4]

\usepackage{url}  %Required
\usepackage{graphicx}  %Required
\def\BibTeX{{\rm B\kern-.05em{\sc i\kern-.025em b}\kern-.08em
    T\kern-.1667em\lower.7ex\hbox{E}\kern-.125emX}}
\begin{document}

\title{Fluid Antenna Enabled Over-the-Air Federated Learning: Joint Optimization of Positioning, Beamforming, and User Selection}

\author{Yang Zhao\textsuperscript{1},~\IEEEauthorblockN{Yue Xiu\textsuperscript{2},~\textup{Minrui Xu}$^{1}$,~\textup{Ping Wang}$^{3}$,~\textup{Dusit Niyato}$^{1}$}
\IEEEauthorblockA{\textsuperscript{1}College of Computing and Data Science,
Nanyang Technological University, Singapore\\
\textsuperscript{2}University of Electronic Science and Technology of China, Chengdu, China \\
\textsuperscript{3}Lassonde School of Engineering, York University, Canada \\
}
}

\author{
\IEEEauthorblockN{Yang Zhao,~\IEEEmembership{Member,~IEEE}, Minrui~Xu,~\IEEEmembership{Member,~IEEE},
Ping~Wang,~\IEEEmembership{Fellow,~IEEE}, \\
Dusit~Niyato,~\IEEEmembership{Fellow,~IEEE}
}
\thanks{Yang Zhao, Minrui Xu, and Dusit Niyato are with Nanyang Technological University, Singapore.
 (Email: {zhao0466@e.ntu.edu.sg, MINRUI001@e.ntu.edu.sg, dniyato@ntu.edu.sg).}
}
\thanks{Ping Wang is with Lassonde School of Engineering, York University, Canada. (Email: \mbox{pingw@yorku.ca}).}

\thanks{A conference version of this work has been accepted by IEEE ICC 2025.}
}

\maketitle

\begin{abstract}

Over-the-air (OTA) federated learning (FL) effectively utilizes communication bandwidth, yet it is vulnerable to errors during analog aggregation. While removing users with unfavorable channel conditions can mitigate these errors, it also reduces the available local training data for FL, which in turn hinders the convergence rate of the training process. To tackle this issue, we propose using fluid antenna (FA) techniques to enhance the degrees of freedom within the channel space, ultimately boosting the convergence speed of FL training. Moreover, we develop a novel approach that effectively coordinates uplink receiver beamforming, user selection, and FA positioning to optimize the convergence rate of OTA FL training in dynamic wireless environments. We address this challenging stochastic optimization by reformulating it as a mixed-integer programming problem by utilizing the training loss upper bound. We then introduce a penalty dual decomposition (PDD) method to solve the mixed-integer mixed programming problem. Experimental results indicate that incorporating FA techniques significantly accelerates the training convergence of FL and greatly surpasses conventional methods.

\end{abstract}


\section{Introduction}

Federated learning (FL) has emerged as a revolutionary distributed machine learning framework that enables collaborative model training across multiple users while preserving data privacy by keeping user data locally~\cite{huang2024federated}. This decentralized approach has gained widespread attention due to its potential to unlock powerful machine learning applications in data-sensitive environments such as healthcare, finance, and the Internet of Things (IoT). A central server aggregates local model updates from participating users in FL systems to improve a shared global model. While effective in principle, practical implementations of FL in wireless environments face significant communication challenges, particularly in settings with many participating users. Traditional orthogonal multiple access methods struggle to handle concurrent transmissions, causing inefficiencies in bandwidth usage and increased communication latency.

To address these limitations, recent advances have proposed using analog signal aggregation for FL. In this method, multiple users can simultaneously transmit their model updates on a shared wireless channel using analog modulation, where the server aggregates the updates automatically through signal superposition~\cite{wen2023survey}. This approach, often referred to as \emph{over-the-air} (OTA) aggregation, takes advantage of the natural waveform superposition property of wireless channels to compute sums of local updates “in the air,” thereby reducing both latency and bandwidth consumption. Consequently, OTA aggregation is a promising solution for large-scale FL deployments~\cite{beitollahi2023federated}. Furthermore, prototype FL systems incorporating wireless aggregation functionalities have demonstrated significant performance improvements~\cite{xu2023accelerating}.

Despite these advances, wireless FL systems remain highly susceptible to noise, propagation loss, and aggregation errors during communication and computation phases~\cite{du2024integrated}. Although OTA aggregation can dramatically improve spectral efficiency, it is notably sensitive to channel fading and interference, forcing transmitters to carefully align phases and powers to avoid destructive interference. The problem is further exacerbated when users experience poor channel conditions, which degrade the overall signal-to-noise ratio (SNR). Stronger users must reduce their transmission power to accommodate weaker users, resulting in a suboptimal communication rate~\cite{kim2023beamforming}. Filtering out users with poor channels could reduce aggregation errors but simultaneously reduce the available training data, ultimately hindering learning performance.

Motivated by these insights, this paper explores a wireless FL framework supported by FA technology and introduces a joint optimization framework designed to enhance the efficiency and scalability of FL in dynamic wireless environments. Specifically, we focus on reducing the global training loss after $T$ communication rounds by designing effective uplink receiver beamforming and user selection strategies~\cite{wong2020performance}. Unlike existing studies, our approach incorporates dynamic channel state information to enable real-time adjustments in user selection, FA positioning, and beamforming strategies for each communication round.

Building upon this foundation, we detail the optimization strategies employed to enhance FL performance in wireless settings. We present a mathematical framework for optimal resource allocation, communication overhead reduction, and accelerated global model convergence. Furthermore, our comprehensive convergence analysis demonstrates how integrating FA technology and advanced optimization techniques significantly improves the robustness and scalability of FL. This work lays the groundwork for robust machine learning deployments in distributed, bandwidth-constrained, and unreliable wireless networks.

The contributions are summarized as follows:
\begin{itemize}
    \item This paper introduces a novel joint optimization framework tailored for over-the-ait FL systems supported by FA technology. To our knowledge, this is the first study to explore the application of FA in FL, leveraging its ability to optimize channel conditions dynamically. Our framework integrates receiver beamforming, FA positioning, and user selection into a cohesive model that maximizes the efficiency of OTA techniques for wireless FL.
    \item We are the first to formulate the optimization problem for FA-assisted FL as a finite-time stochastic optimization, taking advantage of the upper bound of training loss in~\cite{zhang2023joint}. To address this, we propose a novel penalty dual decomposition (PDD) approach that decomposes complex joint optimization into manageable subproblems. This innovative framework ensures computational efficiency and robustness in dynamic wireless FL systems.
    \item We enhance the PDD approach by introducing two key algorithmic components: a greedy algorithm for efficient user selection and the (stochastic successive convex approximation) SCA method to optimize beamforming and FA positioning. These methods enable real-time decision-making, allowing the system to adapt to changing channel conditions while minimizing communication overhead and improving resource allocation.
    \item Through extensive simulations, we demonstrate that FA-assisted FL significantly outperforms traditional methods regarding convergence speed and computational efficiency. Notably, our approach achieves faster convergence rates and reduced complexity, particularly in large-scale user scenarios where conventional systems struggle with communication bottlenecks. These results highlight the transformative potential of integrating FA technology into FL systems.
\end{itemize}

\textbf{Organization:} The rest of this paper is structured as follows. Section~\ref{related-work} reviews existing literature on federated learning and fluid antenna technologies, highlighting developments in OTA aggregation. Section~\ref{system-model} defines the system model, formulates the joint optimization problem under FA constraints, and introduces our PDD and SCA framework. Section~\ref{subsec:channel_bound} addresses the maximum channel gain bound and the resulting convergence implications under \(L\)-smooth and \(\mu\)-strongly convex conditions. Section~\ref{experiments} provides numerical results, validating the proposed method against state-of-the-art benchmarks. Finally, Section~\ref{conclusion} concludes with key findings and future research directions. Table~\ref{tab:notation} summarizes notations used throughout the paper.

\begin{table}[h!]
\centering
\begin{tabular}{|l|p{0.62\linewidth}|}
\hline
\textbf{Symbol} & \textbf{Description} \\ \hline

\(\boldsymbol{a}(\boldsymbol{x}_{u,t}, \boldsymbol{y}_{u,t})\)
    & Array response vector based on the user’s antenna position \\ \hline

\(a_{u,t}\)
    & Per-user transmit amplitude scaling for user \(u\) in round \(t\) \\ \hline

\(\alpha_u\)
    & Auxiliary variable associated with user \(u\) in the PDD formulation \\ \hline

\(\beta_{u,t}\)
    & Path loss coefficient of user \(u\) in round \(t\) \\ \hline

\(c\)
    & Reduction factor for the penalty parameter (\(\kappa(t+1) = c\,\kappa(t)\)) \\ \hline

\(\boldsymbol{e}_t\)
    & Binary vector for user selection in round \(t\) 
      (\(\boldsymbol{e}_t[u] = 1\) if user \(u\) is selected) \\ \hline

\(\eta_{\text{lr}}\)
    & Learning rate for FL model updates \\ \hline

\(\eta_t\)
    & Receive scaling factor for signal normalization at the server in round \(t\) \\ \hline

\(F(\boldsymbol{w})\)
    & Global loss function \\ \hline

\(F_u(\boldsymbol{w}; \mathcal{E}_u)\)
    & Local loss function of user \(u\) \\ \hline

\(\boldsymbol{g}_u\)
    & Gradient vector computed by user \(u\) \\ \hline

\(\boldsymbol{h}_{u,t}\)
    & Uplink channel vector between user \(u\) and server in round \(t\) \\ \hline

\(K\) or \(\bigl|\mathcal{U}_{t}^{s}\bigr|\)
    & Number of active (participating) devices in round \(t\) \\ \hline

\(\kappa\)
    & Penalty parameter in PDD controlling constraint enforcement \\ \hline

\(\lambda\)
    & Weight parameter in constrained optimization \\ \hline

\(L\)
    & Lipschitz constant (smoothness) of the global loss \(F\) \\ \hline

\(\mu\)
    & Strong convexity constant of the global loss \(F\) \\ \hline

\(P_a\)
    & Maximum allowable transmission power (per user) \\ \hline

\(\boldsymbol{q}_t\)
    & Receiver (beamforming) vector at the server in round \(t\) \\ \hline

\(r(\boldsymbol{q}_{t}, \boldsymbol{e}_t)\)
    & Communication penalty function capturing OTA aggregation error \\ \hline

\(S_u\)
    & Number of samples in user \(u\)'s local dataset \\ \hline

\(\sigma_n^2\)
    & Noise variance of the channel \\ \hline

\(\mathcal{U}\)
    & Set of all users \\ \hline

\(u\)
    & Index for a user \\ \hline

\(v_x,\,v_y\)
    & Minimum separation distance for antenna elements in the x and y directions \\ \hline

\(\boldsymbol{w}\)
    & Parameter vector of the global FL model \\ \hline

\(\boldsymbol{x}_{u,t}\)
    & Fluid antenna position of user \(u\) in round \(t\) \\ \hline

\(\boldsymbol{y}_{u,t}\)
    & Fluid antenna vertical position of user \(u\) in round \(t\) \\ \hline

\end{tabular}
\caption{The summary of notations.}
\label{tab:notation}
\end{table}

\section{Related Work} \label{related-work}

\subsection{Federated Learning in Wireless Communication} 
FL has emerged as a promising approach to collaboratively train models across distributed devices without sharing raw data, which is particularly appealing in wireless networks where privacy preservation and limited bandwidth are critical~\cite{mcmahan2017communication, li2020federated}. In FL, participating devices transmit locally computed updates to a central server for aggregation, but doing so over resource-constrained wireless channels raises challenges such as straggler mitigation, communication overhead, and energy efficiency~\cite{konevcny2016federated, tran2019federated, bonawitz2019towards}. Various solutions focus on gradient compression, quantization, and client selection strategies to reduce communication load~\cite{konevcny2016federated, reisizadeh2020fedpaq}, while other studies emphasize joint optimization of FL training and radio resource management to address heterogeneous network conditions and device capabilities~\cite{nishio2019client, chen2021wireless}.

\subsection{Over-the-Air Aggregation for Federated Learning}
Over-the-air aggregation leverages the waveform superposition property of the wireless medium to compute sums of local model updates directly in the channel, thereby reducing latency and alleviating communication bottlenecks for large-scale FL~\cite{zhu2019broadband, elgabli2020harnessing}. By allowing all devices to transmit simultaneously, OTA achieves high spectral efficiency and accelerates training convergence~\cite{zhu2018mimo, yang2020federatedAirComp}, although robust designs must account for imperfect channel state information, synchronization requirements, and noise-induced aggregation errors~\cite{yang2020federatedAirComp, zhang2020gradient}. Existing work addresses these issues through power control, beamforming, and model update compression schemes, establishing OTA computation as a viable method for large-scale FL~\cite{asaad2024joint}.

\subsection{Fluid Antenna in Wireless Communication}
Fluid antennas have gained significant attention for their ability to dynamically optimize wireless channels by adjusting their position, orientation, or shape in response to changing network conditions~\cite{zhang2023fluid, liu2023reconfigurable, wang2024adaptive}. Unlike fixed antennas, fluid antennas utilize liquid metal or mechanical repositioning to enhance spatial diversity, improving signal reception, interference management, and power efficiency~\cite{xu2023fluid, chen2024beamforming}. Recent studies demonstrate that small-scale movements in antenna position can significantly improve line-of-sight (LoS) connectivity and reduce interference in dynamic environments~\cite{xu2023fluid, chen2024beamforming}. Furthermore, the combination of fluid antennas with intelligent reflecting surfaces (IRS) has been explored to maximize coverage and enhance spatial multiplexing in dense networks~\cite{liu2023reconfigurable}. These advancements position fluid antennas as a promising technology for next-generation wireless networks, where adaptive configurations can complement traditional signal processing for enhanced reliability and efficiency~\cite{wang2024adaptive}.

% \textbf{Insights.}  
Unlike existing works that treat antenna positioning and beamforming as separate optimization problems~\cite{zhang2023fluid, liu2023reconfigurable}, our framework jointly optimizes FA positioning, beamforming, and user selection to maximize both communication efficiency and learning performance in wireless FL architectures. By dynamically adapting antenna configurations to channel variations, our approach reduces communication overhead, mitigates interference, and accelerates FL convergence under finite-time stochastic conditions, which solves a challenge that prior studies have only partially addressed. Furthermore, unlike traditional OTA aggregation schemes, which suffer from alignment issues and signal degradation~\cite{xu2023fluid}, our FA-assisted design enhances OTA efficiency by ensuring adaptive spatial diversity and robust model aggregation. This study not only establishes a novel paradigm for integrating FAs into FL systems but also provides a scalable solution for optimizing learning-driven applications in resource-constrained, unreliable networks, positioning it ahead of conventional fixed-antenna and static beamforming strategies~\cite{wang2024adaptive, chen2024beamforming}.

\section{System Model and Problem Formulation} \label{system-model}

\subsection{System Model}

\begin{figure}[h]
  \centering
  \includegraphics[scale=0.35]{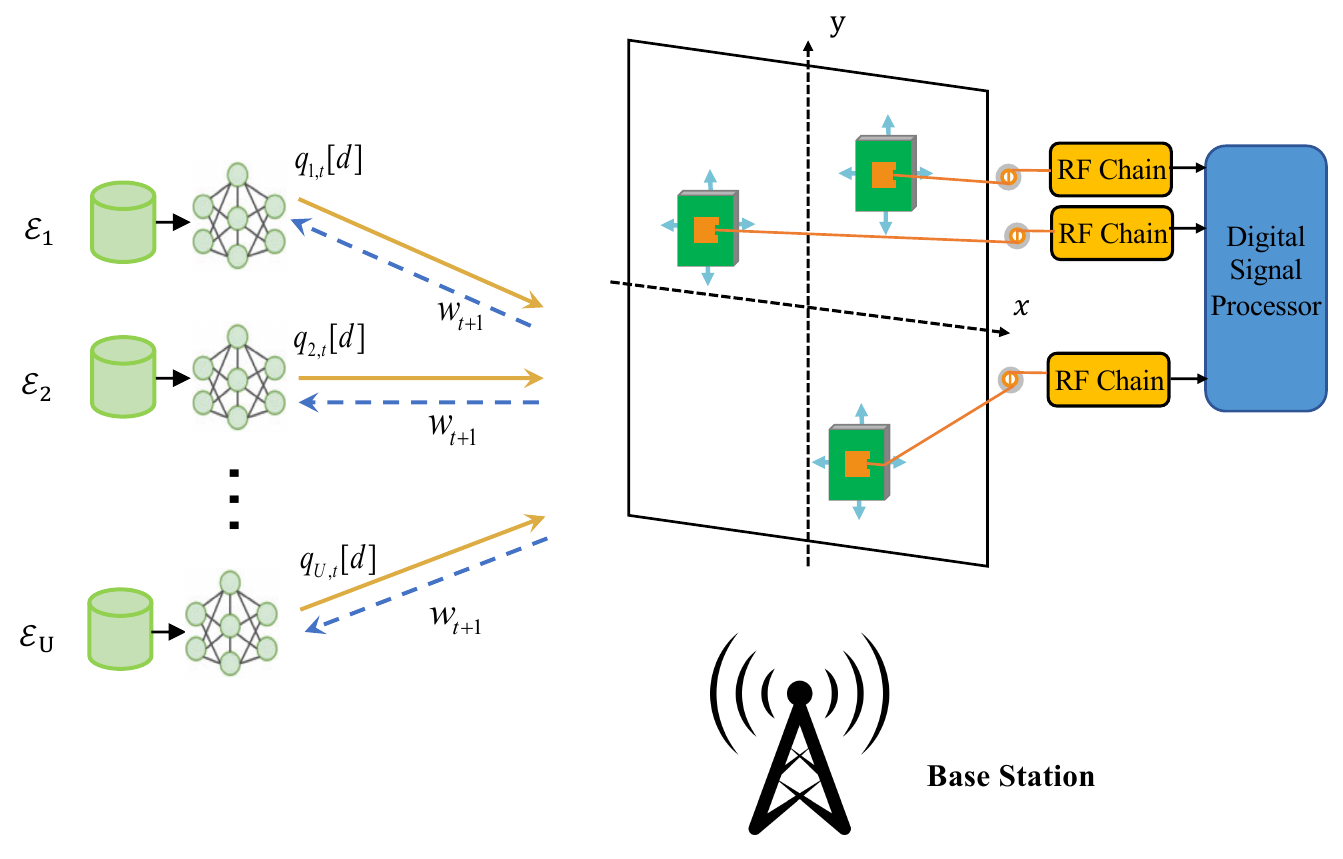}
  \captionsetup{justification=centering}
  \caption{Illustration of the FA-aided FL Systems with OTA.}
\label{FIGUREICC_0}
\end{figure}

We consider a wireless FL system comprising a central server and $U$ local users, as shown in Fig.~\ref{FIGUREICC_0}. The set of users is denoted by $\mathcal{U}=\{1,\ldots, U\}$. Each user 
$u$ holds a local dataset containing $S_{u}$ samples, denoted as 
\begin{align}
    \mathcal{E}_{u}=\{(\boldsymbol{x}_{u,d},y_{u,d}):1\leq d\leq S_{u}\},\label{form1}
\end{align}
where $\boldsymbol{x}_{u,d}\in\mathbb{R}^{b}$ is the feature vector of the $d$-th sample, and $y_{u,d}$ represents its corresponding label. The objective is to collaboratively train a global model hosted on the server, enabling accurate label predictions while preserving the privacy of the user's datasets.

% \textbf{Local and Global Loss Functions.} 
Each user calculates a local loss function on the local dataset, and the empirical loss function for the user $u$ is expressed as follows:
\begin{align}
F_{u}(\boldsymbol{w};\mathcal{E}_{u})=\frac{1}{S_{u}}\sum_{d=1}^{S_{u}}\mathcal{L}(\boldsymbol{w};\boldsymbol{x}_{u,d},y_{u,d}),\label{form2}
\end{align}    
where $\boldsymbol{w}\in\mathbb{R}^{D}$ is the parameter vector of the global model, and $\mathcal{L}$ denotes the loss function for each data sample. 

The training loss for the global model is formulated as a weighted aggregate of the local loss functions from all participating users, which is expressed as
\begin{align}
F(\boldsymbol{w})=\frac{1}{S}\sum_{u=1}^{U}S_{u}F_{u}(\boldsymbol{w};\mathcal{E}_{u}),\label{form3}
\end{align}
where $S=\sum_{u=1}^{U}S_{u}$ represents the total number of training samples from all users. Then, we adopt the Federated Stochastic Gradient Descent (FedSGD) method to minimize the global loss function $F(\boldsymbol{w})$. Each communication round comprises the following steps:
\begin{itemize}
    \item A subset of users is selected for participation denoted as $\mathcal{U}_{t}^{s}$.
    \item The server broadcasts the current global model parameters to the selected users.
    \item Each selected user computes its local gradient based on the global model and transmits the gradient to the server.
    \item The server aggregates the received gradients to update the global model parameters.
\end{itemize}
The detailed procedure for FL communication and model update is described in Algorithm~\ref{alg:FL_round}.

\begin{algorithm}[!ht]
\caption{Federated Learning Communication \& Model Update Algorithm.}
\label{alg:FL_round}
\begin{algorithmic}[1]
\Require 
   Current global model $\boldsymbol{w}_t$, 
   chosen devices $e[u]$, 
   beamformer $\boldsymbol{q}$, 
   antenna positions $\{\boldsymbol{x}_u\}$, 
   local data sets.

\Ensure 
   Updated global model $\boldsymbol{w}_{t+1}$.

\State \textbf{Broadcast} the current model $\boldsymbol{w}_t$ to selected users $\mathcal{U}_{t}^{s}$.
\State \textbf{Local Gradients:} Each user $u$ (for which $e[u]=1$) computes $\boldsymbol{g}_{u,t}$.
\State \textbf{Over-the-Air Aggregation:}
    \Statex \quad Each selected user transmits $f_{u,t}[n] = a_{u,t}\,\frac{g_{u,t}[n]}{v_{u,t}}$ 
    \Statex \quad under power constraints $|a_{u,t}|^2 \le P_a$.
    \State \textbf{Server receives} $\boldsymbol{y}_{n,t} = \sum_{u\in \mathcal{U}_{t}^{s}}\boldsymbol{h}_{u,t}\,f_{u,t}[n] + \boldsymbol{s}_{n,t}$.
    \State \textbf{Apply beamforming} $\boldsymbol{q}$ to obtain scaled sum 
    $\tilde{y}_t[n] = \frac{\boldsymbol{q}^H \boldsymbol{y}_{n,t}}{\sqrt{\eta_t}}$.

\State \textbf{Update Global Model:}
    \[
      \boldsymbol{g}_t 
      = \sum\nolimits_{u \in U} a_{u,t}\,\frac{\boldsymbol{g}_{u,t}}{v_{u,t}},
      \quad
      \boldsymbol{w}_{t+1} 
      \;\leftarrow\; \boldsymbol{w}_t - \eta\,\boldsymbol{g}_t.
    \]

\State \textbf{Return} $\boldsymbol{w}_{t+1}$.
\end{algorithmic}
\end{algorithm}

The central server is equipped with a $2D$ phase array with $N_{T}$ FAs, while each user has a single antenna. The uplink LoS channel between user $u$ and the server during communication round $t$ is represented by $\boldsymbol{h}_{u,t}\in\mathbb{C}^{N_{T}}$, and it is given by
$\boldsymbol{h}_{u,t}=\beta_{u,t}\boldsymbol{a}(\boldsymbol{x}_{u,t},\boldsymbol{y}_{u,t})$,
where $\beta_{u,t}$ is the complex path gain, and 
\begin{align}
    &\boldsymbol{a}(\boldsymbol{x}_{u,t},\boldsymbol{y}_{u,t})=[e^{j\frac{2\pi}{\lambda}\rho_{u,t,1}},e^{j\frac{2\pi}{\lambda}\rho_{u,t,2}}, \ldots,e^{j\frac{2\pi}{\lambda} \rho_{u,t,N_{T}}}]^{T},\label{form4}
\end{align}
where $\boldsymbol{a}(\boldsymbol{x}_{u,t},\boldsymbol{y}_{u,t})\in\mathbb{C}^{N_{T}\times 1}$, $\rho_{u,t,i}=x_{u,t,i}\cos(\theta_{u,t})+y_{u,t,i}\sin(\phi_{u,t})$, $\forall~i\in\{1,\cdots,N_{T}\}$, and $\boldsymbol{x}_{u,t}=[x_{u,t,1}$,$x_{u,t,2}$,\\
$\ldots$,$x_{u,t,N_{T}}]^{T}\in\mathbb{C}^{N_{T}\times 1}$ and $\boldsymbol{y}_{u,t}=[y_{u,t,1}$,$y_{u,t,2}$,$\ldots$,$y_{u,t,N_{T}}]^{T}$\\
$\in\mathbb{C}^{N_{T}\times 1}$ denote the FA position coordinates, $\theta_{u,t}$ and $\phi_{u,t}$ are the azimuth angle of arrival (AoA) and the elevation AoA of the line-of-sight (LoS) channel, respectively. 

To improve communication efficiency, we employ OTA analog aggregation through a multiple access channel, enabling the efficient combination of local gradients. During each communication round, as shown in Fig.~\ref{FIGUREICC_0}, users transmit their local gradients simultaneously over the same frequency band, which results in a weighted sum of the transmitted signals being received by the server. In the $t$-th communication round, each selected user employs $D$ symbol durations to transmit its local gradient. The local gradient vector $\boldsymbol{g}_{u,t}=[g_{u,t}[1],\cdots,g_{u,t}[N]]$ for each user $u$ is first normalized by a scalar $v_{u,t}$ and then scaled by a transmission weight $a_{u,t}$ before transmission. The transmitted signal for the $n$-th component of the local gradient, denoted as $f_{u,t}[n]$, is expressed as
\begin{align}
f_{u,t}[n] = a_{u,t} \frac{g_{u,t}[n]}{v_{u,t}}.\label{form5}
\end{align}
To support processing at the server, each chosen user $u$ transmits its local gradient normalization factor
\begin{align}
v_{u,t} = \frac{\|\boldsymbol{g}_{u,t}\|}{\sqrt{D}},\label{form6}
\end{align}
via the uplink signaling channel. This channel is assumed to be a dedicated digital link with perfect signal reception. Let 
$\boldsymbol{f}_{u,t} = [f_{u,t}[1], \ldots, f_{u,t}[D]]^{T}$
represent the signal vector corresponding to the local gradient $\boldsymbol{g}_{u,t}$. From  (\ref{form6}), the average transmission power used for each component of $\boldsymbol{g}_{u,t}$ by user $u$ in round $t$ is given by
\begin{align}
\frac{\|\boldsymbol{f}_{u,t}\|^{2}}{D} = |a_{u,t}|^{2}.\label{form7}
\end{align}
To ensure that users adhere to power constraints, the transmission power is limited by a maximum allowable average power $P_a$, such that
\begin{align}
|a_{u,t}|^{2} \leq P_a, \quad \forall u, t.\label{form8}
\end{align}

The received signal at the server for the $n$-th symbol interval during communication round $t$ can be written as
\begin{align}
\boldsymbol{y}_{n,t}=\sum\nolimits_{u\in\mathcal{U}_{t}^{s}}\boldsymbol{h}_{u,t}f_{u,t}[n]+\boldsymbol{s}_{n,t},\label{form9}
\end{align}
where $\boldsymbol{s}_{n,t}\sim\mathcal{CN}(\boldsymbol{0},\sigma_{n}^{2}\boldsymbol{I})$ represents the additive white Gaussian noise (AWGN) at the receiver for the 
$n$-th channel use, which is independent and identically distributed (i.i.d.) over time $t$. The server applies receive beamforming to process the incoming signal. Let $\boldsymbol{q}_{t}\in\mathbb{C}^{N}$ denote the received beamforming vector during the communication round $t$, subject to the constraint $\|\boldsymbol{q}_{t}\|_{2}=1$, and $\eta_{t}\in\mathbb{R}^{+}$ denotes the scaling factor for reception. Thus, the processed received signal is 
$\tilde{y}_{t}[n]=\frac{\boldsymbol{q}_{t}^{H}\boldsymbol{y}_{n,t}}{\sqrt{\eta_{t}}}.$

\subsection{Problem Formulation}

To enhance the learning efficiency of FL, we aim to improve the training convergence rate by minimizing the expected global loss function after $T$ communication rounds. This is achieved through joint optimization of the selection of devices $\mathcal{U}_{t}^{s}$, FA positioning, the transmit weights $\{a_{u,t}\}$ of the devices, and the server-side processing, including the beamforming vector  $\boldsymbol{q}_{t}$ and the scaling factor $\eta_{t}$. The optimization problem can be expressed as
\begin{subequations}
    \begin{align} \min_{\{\boldsymbol{x}_{u,t},\boldsymbol{y}_{u,t},\boldsymbol{q}_{t},\boldsymbol{e}_{t},\eta_{t},\{a_{u,t}\}\}_{t=0}^{T-1}}&\mathbb{E}[F(\boldsymbol{w}_{T})],\label{form10a}\\
    \text{s.t.}~ 
    &|a_{u,t}|^{2}\leq P_{a},&\label{form10b}\\
    &\|\boldsymbol{q}_{t}\|_{2}=1,&\label{form10c}\\ 
    &\eta_{t}>0,&\label{form10d}\\ 
    &\boldsymbol{e}_{t}\in\{0,1\}^{U},&\label{form10e}\\
    &|x_{u,t,n_{1}}-x_{u,t,n_{2}}|\geq v_{x},~n_{1}\neq n_{2},&\nonumber\\
    &|y_{u,t,n_{1}}-y_{u,t,n_{2}}|\geq v_{y},~n_{1}\neq n_{2},&\label{form10f}\\
    &x_{u,t,n_{1}},x_{u,t,n_{2}}\in\mathcal{C}_{x},&\nonumber\\
    &y_{u,t,n_{1}},y_{u,t,n_{2}}\in\mathcal{C}_{y}, &\label{form10g}
    \end{align}\label{form10}%
\end{subequations}%
where $\mathcal{C}_{x}=[x_{u,t}^{\min},x_{u,t}^{\max}]$ and $\mathcal{C}_{y}=[y_{u,t}^{\min},y_{u,t}^{\max}]$ are the fluid region of FAs, and $x_{u,t}^{\min}/x_{u,t}^{\max}$ and $y_{u,t}^{\min}/y_{u,t}^{\max}$ are  minimization/maximum coordinates. $\mathbb{E}$ denotes the expectation calculated over the receiver's noise, and $\boldsymbol{e}_{t}$ is a binary vector that indicates device selection for round $t$. Specifically, if the $u$-th element $\boldsymbol{e}_{t}[u]=1$, it signifies that device $u$ is selected to contribute to model updating for communication round $t$; otherwise, $\boldsymbol{e}_{t}[u]=0$ indicates it has not been selected. Notably, $\boldsymbol{e}_{t}$ and $\mathcal{U}_{t}^{s}$ are equivalent, with $\mathcal{U}_{t}^{s}=\{u:\boldsymbol{e}_{t}[u]=1,u\in\mathcal{U}\}$. Unlike traditional channels with FPA, the FA channel represented by $\boldsymbol{h}_{u,t}$ depends on the FA's position. This position affects the channel matrix and the global loss function $\mathbb{E}[F(\boldsymbol{w}_{T})]$. Therefore, the FA location constraint is incorporated into the optimization problem (\ref{form10}), where $v_{x}$ and $v_{y}$ represent the minimum separation distance between antennas, and $\mathcal{C}$ denotes the movement region of the FAs.

The upper bound is derived based on specific assumptions about the global loss function 
$F(\boldsymbol{w})$, commonly encountered in the stochastic optimization literature \cite{beitollahi2023federated}. Building on these assumptions and following the derivation outlined in \cite{wang2023distributed}, the expected difference between the global loss function at round $(t+1)$ and the optimal loss can be bounded as
\begin{align}
&\mathbb{E}[F(\boldsymbol{w}_{t+1})-F(\boldsymbol{w}^{*})] \nonumber \\
&\quad \leq\phi_{t}\mathbb{E}[F(\boldsymbol{w}_{t})-F(\boldsymbol{w}^{*})]+ \frac{\alpha_{1}}{L}r(\boldsymbol{q}_{t},\boldsymbol{e}_{t};\mathcal{T}_{t}),\label{form11}
\end{align}
where $\phi_{t}=1-\frac{\mu}{L}(1-2\alpha_{2}r(\boldsymbol{q}_{t},\boldsymbol{e}_{t};\mathcal{T}_{t}))$, $\mathcal{T}_{t}=\{\boldsymbol{h}_{u,t}:u\in\mathcal{U}\}$, and
\begin{align}
&r(\boldsymbol{q}_{t},\boldsymbol{e}_{t};\mathcal{T}_{t})=\frac{4}{U^{2}}\big (\sum\nolimits_{u=1}^{U}(1-\boldsymbol{e}_{t}[u])S_{u} \big )^{2}+\nonumber\\
&\frac{\sigma_{n}^{2}}{P_{a}(\sum\nolimits_{u=1}^{U}\boldsymbol{e}_{t}[u]S_{u})^{2}}\max\nolimits_{1\leq u\leq U}\frac{\boldsymbol{e}_{t}[u]S_{u}^{2}}{|\boldsymbol{q}_{t}^{H}\boldsymbol{h}_{u,t}(\boldsymbol{x}_{u,t},\boldsymbol{y}_{u,t})|^{2}},\label{form12}
\end{align}
where $\mu\geq 0$ is strongly convex constant, $L\geq 0$ is Lipschitz constant, and $\alpha_{1}\geq 0$, $\alpha_{2}\geq 1$ are coefficients from the loss function.

Let $\boldsymbol{w}_{0}$ be the initial model parameter vector. Applying the result from \cite{liu2021reconfigurable},
the upper bound on the expected global loss after $T$ communication rounds is given by $\mathbb{E}[F(\boldsymbol{w}_{t+1})-F(\boldsymbol{w}^{*})]$ for $t=0,\ldots, T-1$, and we have the following upper bound after $T$ communication rounds
\begin{align}
&\mathbb{E}[F(\boldsymbol{w}_{T})-F(\boldsymbol{w}^{*})]\leq(\prod_{t=0}^{T-1}\phi_{t})\mathbb{E}[F(\boldsymbol{w}_{0})-F(\boldsymbol{w}^{*})]+\frac{\alpha_{1}}{L}(\nonumber\\
&
\sum_{t=0}^{T-2}(\prod_{\tau=t+1}^{T-1}\phi_{r})r(\boldsymbol{q}_{t},\boldsymbol{e}_{t};\mathcal{T}_{t}))+r \big (\boldsymbol{q}_{T-1},\boldsymbol{e}_{T-1};\mathcal{T}_{T-1}) \big ).\label{form13}
\end{align}

It is important to note that minimizing 
$\mathbb{E}[F(\boldsymbol{w}_{T})]$ in (\ref{form10a}) is equivalent to minimizing 
$\mathbb{E}[F(\boldsymbol{w}_{T})-F(\boldsymbol{w}^{*})]$. However, directly optimizing 
$\mathbb{E}[F(\boldsymbol{w}_{T})-F(\boldsymbol{w}^{*})]$ is challenging. Instead, we focus on minimizing its upper bound as shown in (\ref{form13}). Since 
$\phi_{t}$
is a growing function of 
$r(\boldsymbol{q}_{t},\boldsymbol{e}_{t};\mathcal{T}_{t}))$, the upper bound in (\ref{form13}) also increases with 
$r(\boldsymbol{q}_{t},\boldsymbol{e}_{t};\mathcal{T}_{t}))$, $t = 0,\ldots,T-1$. Therefore, to reduce the upper bound, it suffices to minimize 
$r(\boldsymbol{q}_{t},\boldsymbol{e}_{t};\mathcal{T}_{t}))$ at each round with respect to $(\boldsymbol{q}_{t},\boldsymbol{e}_{t})$.

This combined optimization problem for user selection and receiver beamforming is formulated as in (\ref{form14}), where the subscript $t$ is omitted for simplicity.
\begin{subequations}
\begin{align} \min_{\mathcal{U}}&\frac{4}{U^{2}}(\sum\nolimits_{u=1}^{U}(1-e[u])S_{u})^{2}+\nonumber\\
&\frac{\sigma_{n}^{2}}{P_{a}(\sum\nolimits_{u=1}^{U}e[u]S_{u})^{2}}\max\nolimits_{1\leq u\leq U}\frac{e[u]S_{u}^{2}}{|\boldsymbol{q}^{H}\boldsymbol{h}_{u}(\boldsymbol{x}_{u},\boldsymbol{y}_{u})|^{2}},\label{form14a}\\
\text{s.t.}~
&(\ref{form10c}),(\ref{form10e}),(\ref{form10f}),(\ref{form10g}),\label{form14b}
\end{align}\label{form14}%
\end{subequations}
where $\mathcal{U}=\{\boldsymbol{x}_{u},\boldsymbol{y}_{u},\boldsymbol{q},\boldsymbol{e}\}$.

Due to the non-convex nature of the problem (\ref{form14}), directly solving it may lead to suboptimal solutions. To address this challenge, we introduce an auxiliary variable $c$, which helps reformulate the complex non-convex optimization into a more tractable form, and we let
\begin{align}
c=\max\nolimits_{1\leq u\leq U}\frac{e[u]S_{u}^{2}}{|\boldsymbol{q}^{H}\boldsymbol{h}_{u}(\boldsymbol{x}_{u},\boldsymbol{y}_{u})|^{2}}.\label{form15}
\end{align}
According to (\ref{form15}), we have $(e[u]S_{u}^{2})/|\boldsymbol{q}^{H}\boldsymbol{h}_{u}(\boldsymbol{x}_{u},\boldsymbol{y}_{u}))|^{2}\leq c$. Because of the complexity of the problem and following the framework of the PDD algorithm, it is essential to introduce auxiliary variables $\bar{e}[u]$, $\tilde{e}[u]$, $\hat{e}[u]$, and we have
$e[u]=\tilde{e}[u]=\bar{e}[u]=\hat{e}[u],$ and (\ref{form10e}) is equivalently written as
$e[u](1-e[u])=0$, where $ 0\leq e[u]\leq 1.$ Then, we further introduce the auxiliary variables $\alpha_{u}$, $\tilde{\boldsymbol{x}}_{u,n_{1}}$ and $\tilde{\boldsymbol{y}}_{u,n_{1}}$, and we let $\alpha_{u}=|\boldsymbol{q}^{H}\boldsymbol{h}_{u}(\boldsymbol{x}_{u},\boldsymbol{y}_{u})|^{2}$ and $\tilde{\boldsymbol{x}}_{u,n_{1}}=\boldsymbol{x}_{u,n{1}}-\boldsymbol{x}_{u,n{2}}$ and $\tilde{\boldsymbol{y}}_{u,n_{1}}=\boldsymbol{y}_{u,n{1}}-\boldsymbol{y}_{u,n{2}}$. Therefore, the problem is rewritten as
\begin{subequations}
\begin{align} \min_{\mathcal{U}}&\frac{4}{U^{2}}(\sum\nolimits_{u=1}^{U}(1-\tilde{e}[u])S_{u})^{2}+\frac{\sigma_{n}^{2}c}{P_{a}(\sum\nolimits_{u=1}^{U}e[u]S_{u})^{2}},\label{form16a}\\
\text{s.t.}~ 
&e[u]=\tilde{e}[u]=\hat{e}[u]=\bar{e}[u],~\forall~u,&\label{form16b}\\
&\frac{e[u]S_{u}^{2}}{\alpha_{u}}\leq c,&\label{form16c}\\
&\alpha_{u}=|\boldsymbol{q}^{H}\boldsymbol{h}_{u}(\boldsymbol{x}_{u},\boldsymbol{y}_{u})|^{2},&\label{form16d}\\
&e[u](1-e[u])=0, 0\leq e[u]\leq 1,&\label{form16e}\\
&\boldsymbol{x}_{u}\in\mathcal{C}_{x},\boldsymbol{y}_{u}\in\mathcal{C}_{y}&\label{form16f}\\
&|\tilde{\boldsymbol{x}}_{u,n_{1}}|_{2}\geq v_{x},|\tilde{\boldsymbol{y}}_{u,n_{1}}|_{2}\geq v_{y},~n_{1}\neq n_{2},&\label{form16g}\\
&\tilde{\boldsymbol{x}}_{u,n_{1}}=\boldsymbol{x}_{u,n{1}}-\boldsymbol{x}_{u,n{2}},\tilde{\boldsymbol{y}}_{u,n_{1}}=\boldsymbol{y}_{u,n{1}}-\boldsymbol{y}_{u,n{2}}.&\label{form16h}
\end{align}\label{form16}%
\end{subequations}

For the non-convex constraint in (\ref{form16f}), we use SCA to cope with constraint (\ref{form16f}) and $\boldsymbol{x}_{u,t,n_{2}}$ is given as initial FA positions. Constraint (\ref{form16g}) can be rewritten in the following form
\begin{align}
&\|\tilde{\boldsymbol{x}}_{u,n_{1}}^{(i-1)}\|_{2}-2(\tilde{\boldsymbol{x}}_{u,n_{1}}^{(i-1)})^{T}\tilde{\boldsymbol{x}}_{u,n_{1}}\geq v_{x},\nonumber\\
&\|\tilde{\boldsymbol{y}}_{u,n_{1}}^{(i-1)}\|_{2}-2(\tilde{\boldsymbol{y}}_{u,n_{1}}^{(i-1)})^{T}\tilde{\boldsymbol{y}}_{u,n_{1}}\geq v_{y},\label{form17}
\end{align}
where $\tilde{\boldsymbol{x}}_{u,n_{1}}^{(i-1)}=\boldsymbol{x}_{u,n_{2}}$, $\tilde{\boldsymbol{y}}_{u,n_{1}}^{(i-1)}=\boldsymbol{y}_{u,n_{2}}$. 
We let $\boldsymbol{a}(\boldsymbol{x}_{u},\boldsymbol{y}_{u})=\boldsymbol{b}_{u}$, and $\boldsymbol{b}_{u}$ is an auxiliary variable introduced to decouple the nonlinear relationship between the antenna positions and the array response matrix. Given that all elements in the matrix response must satisfy
the constant modulus constraint, we introduce the constant modulus constraint for $|\boldsymbol{b}_{u}(i)|=1$. Then, we introduce the auxiliary variable $\gamma_{u}$, and we have
\begin{align}
\gamma_{u}=\boldsymbol{q}^{H}\boldsymbol{b}_{u}\beta_{u}.\label{form18}
\end{align}
Additionally, we continue to introduce the auxiliary variables $\eta$, $\tilde{\eta}$, $\hat{\eta}$, $\bar{\eta}$, $\tilde{\boldsymbol{u}}$, $\tilde{c}$ and let $\eta=\frac{c\sigma_{n}^{2}}{P_{a}(\sum\nolimits_{u=1}^{U}e[u]S_{u})^{2}}$, $\tilde{\eta}=\sum\nolimits_{u=1}^{U}e[u]S_{u}$,
$\hat{\eta}=\sum\nolimits_{u=1}^{U}\tilde{e}[u]S_{u}$, $\boldsymbol{u}=\tilde{\boldsymbol{u}}$, $\tilde{c}=\bar{\eta}\tilde{\eta}$. Thus, problem (\ref{form16}) can be further expressed as
\begin{subequations}
\begin{align} \min_{\bar{\mathcal{U}}}&\frac{4}{U^{2}}(U-\hat{\eta})^{2}+\eta,\label{form19a}\\
\text{s.t.}~ 
&\bar{e}[u]S_{u}^{2}\leq\hat{\alpha}_{u},\alpha_{u}=|\gamma_{u}|^{2}, P_{a}\eta\tilde{c}= c\sigma_{n}^{2},&\label{form19b}\\
&\tilde{\eta}=\sum\nolimits_{u=1}^{U}\hat{e}[u]S_{u},\hat{\eta}=\sum\nolimits_{u=1}^{U}\tilde{e}[u]S_{u},&\label{form19c}\\
&|\boldsymbol{b}_{u}(i)|=1,&\label{form19d}\\
&\tilde{\alpha}_{u}=\alpha_{u},\tilde{c}=c, \hat{\alpha}_{u}=\tilde{\alpha}_{u}\tilde{c}, \gamma_{u}=\boldsymbol{q}^{H}\boldsymbol{b}_{u}\beta_{u}, \tilde{\eta}=\hat{\eta}, &\nonumber\\
&\bar{\eta}=\hat{\eta}\tilde{\eta}, \boldsymbol{a}(\boldsymbol{x}_{u},\boldsymbol{y}_{u})=\boldsymbol{b}_{u}, \boldsymbol{u}=\tilde{\boldsymbol{u}},&\label{form19e}\\
&(\ref{form16b}),(\ref{form16e}),(\ref{form16f}),(\ref{form16g}),(\ref{form16h}),(\ref{form17}),&\label{form19f}
\end{align}\label{form19}%
\end{subequations}
where $\bar{\mathcal{U}}=\{e[u],\tilde{e}[u],\bar{e}[u],\hat{e}[u],\alpha_{u},\tilde{\alpha}_{u},\hat{\alpha}_{u},\gamma_{k},\eta,\bar{\eta},\hat{\eta},\tilde{\eta},c,\tilde{c},$ $\boldsymbol{u},\tilde{\boldsymbol{u}},\boldsymbol{b}_{u},\boldsymbol{x}_{u,n_{1}},\tilde{\boldsymbol{x}}_{u,n_{1}},\boldsymbol{y}_{u,n_{1}},\tilde{\boldsymbol{y}}_{u,n_{1}}\}$.

\section{Proposed PDD-Based FA-FL Algorithm}

Following the execution framework of the PDD-based algorithm, the penalty parameter $\kappa$ is updated using the formula
$\kappa(\tilde{t}+1) = c\kappa(\tilde{t})$, $(0<c<1)$, depending on the level of constraint violation, and the dual variables are adjusted based on (\ref{form21}) at the top of the next page, where $\tilde{t}$ is the number of outer iterations. Following the approach outlined in \cite{shi2020penalty}, the flow of the proposed PDD-based algorithm is summarized in Algorithm~\ref{alg:PDD}. The dual variables are given by
\begin{align}
&\lambda_{\mathcal{U}_{1}}^{(\tilde{t}+1)}=\lambda_{\mathcal{U}_{1}}^{(\tilde{t})}+\frac{1}{\kappa^{(\tilde{t})}}\chi,\boldsymbol{\lambda}_{\boldsymbol{
\mathcal{U}}_{1}}^{(\tilde{t}+1)}=\boldsymbol{\lambda}_{\boldsymbol{
\mathcal{U}}_{1}}^{(\tilde{t}+1)}+\frac{1}{\kappa^{(\tilde{t})}}\boldsymbol{\chi},\nonumber\\
&\|\boldsymbol{h}(\mathcal{U}_{1}^{(\tilde{t})})\|_{\infty}=\max\{|\mathcal{U}_{1}|\},\|\boldsymbol{h}(\boldsymbol{\mathcal{U}}_{1}^{(\tilde{t})})\|_{\infty}=\max\{|\boldsymbol{\mathcal{U}}_{1}|\},\label{form20}%
\end{align}
where $t$ is the number of outer iterations, $\mathcal{U}_{1}\in\{(e[u]-\tilde{e}[u]),(e[u]-\hat{e}[u]),(e[u]-\bar{e}[u]),(\gamma_{u}-\boldsymbol{q}^{H}\boldsymbol{b}_{u}\beta_{u}),(\alpha_{u}-\tilde{\alpha}_{u}),(c-\tilde{c}),(\hat{\alpha}_{u}-\tilde{\alpha}_{u}\tilde{c}),(\tilde{\eta}-\hat{\eta}),(\bar{\eta}-\hat{\eta}\tilde{\eta})\},$
and $\boldsymbol{\mathcal{U}}_{1}\in\{(\boldsymbol{a}(\boldsymbol{x}_{u},\boldsymbol{y}_{u})-\boldsymbol{b}_{u}),(\boldsymbol{q}-\tilde{\boldsymbol{q}})\}$. $\mathcal{U}_{1}$ and $\boldsymbol{\mathcal{U}}_{1}$ list the sets of scalar and vector equality constraints, respectively, whose residuals are penalized and corrected through the dual variable updates.

\begin{figure*}
\begin{subequations}
\begin{align} 
\min_{\bar{\mathcal{U}}}&\frac{4}{K^{2}}(M-\hat{\eta})^{2}+\eta+\frac{1}{2\kappa}\big (\big (\sum\nolimits_{u=1}^{U}|e[u]-\tilde{e}[u]+\kappa\lambda_{\tilde{e}_{u}}|^{2}+|e[u]-\hat{e}[u]+\kappa\lambda_{\hat{e}_{u}}|^{2}+|e[u]-\bar{e}[u]+\kappa\lambda_{\bar{e}_{u}}|^{2}\nonumber\\
&+|\gamma_{u}-\boldsymbol{q}^{H}\boldsymbol{b}_{u}\beta_{u}+\kappa\lambda_{\gamma_{u}}|^{2}+\|\boldsymbol{a}(\boldsymbol{x}_{u},\boldsymbol{y}_{u})-\boldsymbol{b}_{u}+\kappa\lambda_{\boldsymbol{b}_{u}}\|^{2}+\|\alpha_{u}-\tilde{\alpha}_{u}+\kappa\lambda_{\tilde{\alpha}_{u}}\|^{2}+\|\hat{\alpha}_{u}-\tilde{\alpha}_{u}\tilde{c}+\kappa\lambda_{\hat{\alpha}_{u}}\|^{2}\nonumber\\
&+\|\tilde{\boldsymbol{x}}_{u,n_{1}}-(\boldsymbol{x}_{u,n_{1}}-\boldsymbol{x}_{u,n_{2}})+\kappa\boldsymbol{\lambda}_{\tilde{\boldsymbol{x}}_{u,n_{1}}}\|^{2}+\|\tilde{\boldsymbol{y}}_{u,n_{1}}-(\boldsymbol{y}_{u,n_{1}}-\boldsymbol{y}_{u,n_{2}})+\kappa\boldsymbol{\lambda}_{\tilde{\boldsymbol{y}}_{u,n_{1}}}\|^{2}\big )+\|c-\tilde{c}+\kappa\lambda_{\tilde{c}}\|^{2}\nonumber\\
&+|\tilde{\eta}-\hat{\eta}+\kappa\lambda_{\tilde{\eta}}|^{2}+|\bar{\eta}-\hat{\eta}\tilde{\eta}+\kappa\lambda_{\bar{\eta}}|^{2}+\|\boldsymbol{q}-\tilde{\boldsymbol{q}}+\kappa\boldsymbol{\lambda}_{\boldsymbol{q}}\|^{2} \big ),\label{form21a}\\
\text{s.t.}~ 
&(\ref{form16e}), (\ref{form16f}), (\ref{form16g}), (\ref{form16h}),(\ref{form17}), (\ref{form19b}),(\ref{form19c}),(\ref{form19d}). &\label{form21b}
\end{align}\label{form21}%
\end{subequations}
\hrulefill
\end{figure*}

The proposed algorithm’s inner loop organizes the optimized variables into three blocks, which are updated sequentially. However, the variables within each block can be updated in parallel during each optimization round. Once the inner loop meets the specified accuracy tolerance, the algorithm progresses to the outer loop, where it updates the dual variables and penalty parameters until convergence to a stationary solution set is achieved. A detailed explanation of the update procedures is as follows.

\begin{algorithm}[t]
\caption{PDD-Based FA-FL Algorithm for Problem~(\ref{form15})}
\label{alg:PDD}
\begin{algorithmic}[1]
\Require
   Initial variable set $\bar{\mathcal{U}}^{(0)}$, 
   penalty parameter $\kappa^{(0)} > 0$, 
   dual variables $\boldsymbol{\lambda}^{(0)}$, 
   reduction factor $0 < c < 1$, 
   maximum outer iterations $T_{\mathrm{max}}$.

\Statex \textbf{Output:} Optimized solution set $\bar{\mathcal{U}}$.

\State $\tilde{t} \leftarrow 0$ \quad \Comment{Outer iteration index}

\Repeat  \Comment{(Outer loop begins)}

  \State \textbf{Inner Loop:} 
  \Statex \quad \textbf{(a) First Round Update:}
  \Statex \quad\quad 
   \(\{e[u]\}\) is computed via \eqref{form23}.
  \Statex \quad\quad 
   \(\{\alpha_{u}, c, \hat{\alpha}_{u}, \gamma_{u}, \tilde{\eta}, 
        \tilde{\boldsymbol{x}}_{u,n_1}, \tilde{\boldsymbol{y}}_{u,n_1}\}\)
   is updated according to \eqref{form25}.
  \Statex \quad\quad
   \(\{\tilde{\boldsymbol{q}}\}\) is computed via \eqref{form27}.
  \Statex \quad\quad
   \(\{\boldsymbol{b}_{u}\}\) is computed via \eqref{form28}.

  \Statex \quad \textbf{(b) Second Round Update:}
  \Statex \quad\quad
   \(\{\tilde{e}[u], \hat{e}[u], \bar{e}[u]\}\) is computed using \eqref{form29}.
  \Statex \quad\quad
   \(\{\tilde{\alpha}_{u}, \hat{\eta}, \eta, \boldsymbol{q}\}\) is computed 
   based on \eqref{form31}.

  \Statex \quad \textbf{(c) Third Round Update:}
  \Statex \quad\quad
   \(\{\boldsymbol{x}_{u,n_1}, \boldsymbol{y}_{u,n_1}\}\) is computed 
   using \eqref{form35}.
  \Statex \quad\quad
   \(\{\bar{\eta}, \tilde{c}\}\) is computed via \eqref{form36}.

  \Statex \quad \textbf{Check Inner-Loop Convergence:}
  \Statex \quad\quad 
  If $\|\boldsymbol{h}(\mathcal{X}^{(\tilde{t})})\|_{\infty} \le \epsilon_{\mathrm{inner}}^{(\tilde{t})}$, 
  \textbf{break} the inner loop; otherwise, repeat (a)--(c).
  \State \textbf{Outer Loop Update:}  
  \Statex \quad 
  \textbf{If} $\|\boldsymbol{h}(\mathcal{X}^{(\tilde{t})})\|_{\infty} \le \epsilon_{\mathrm{inner}}^{(\tilde{t})}$:
  \Statex \quad \quad Update $\boldsymbol{\lambda}^{(\tilde{t}+1)}$ via (\ref{form20}), $\kappa^{(\tilde{t}+1)} = \epsilon^{(\tilde{t})}$,
  \Statex \quad \textbf{Else}:
  \Statex \quad \quad $\boldsymbol{\lambda}^{(\tilde{t}+1)} \leftarrow \boldsymbol{\lambda}^{(\tilde{t})}$,
                       $\kappa^{(\tilde{t}+1)} \leftarrow \kappa^{(\tilde{t})}$.

  \Statex \quad $\tilde{t} \leftarrow \tilde{t} + 1$ 
  \State \textbf{Check Outer-Loop Termination:}
  \Statex \quad 
   If $\tilde{t} \ge T_{\mathrm{max}}$ \textbf{ or } 
   $\|\boldsymbol{h}(\mathcal{X}^{(\tilde{t})})\|_\infty \le \epsilon_{\mathrm{outer}}$, 
   then \textbf{stop}.
\Until \text{stopping condition is met}
\State \textbf{return} \(\bar{\mathcal{U}}\)\quad \Comment{Set of final optimized solutions}
\end{algorithmic}
\end{algorithm}

\subsection{1-st round of optimization}
The optimization problem of $e[u]$ is given by
\begin{subequations}
\begin{align} 
\min_{e[u]}&|e[u]-\tilde{e}[u]+\kappa\lambda_{\tilde{e}_{u}}|^{2}+|e[u]-\hat{e}[u]+\kappa\lambda_{\hat{e}_{u}}|^{2}+\nonumber\\
&|e[u]-\bar{e}[u]+\kappa\lambda_{\bar{e}_{u}}|^{2},\label{form22a}\\
\text{s.t.}~ 
&(1-e[u])e[u]=0, 0\leq e[u]\leq 1.&\label{form22b}
\end{align}\label{form22}%    
\end{subequations}
By introducing the Lagrange multiplier $\lambda_{1,u}$
to constraint (\ref{form22b}), the optimal $e[u]$ can be expressed as
\begin{align}
e[u]=\left\{\begin{matrix}
\frac{\lambda_{1,u}/2+\tilde{e}[u]+\hat{e}[u]+\bar{e}[u]-\kappa(\lambda_{\tilde{e}[u]}+\lambda_{\tilde{e}[u]}+\lambda_{\bar{e}[u]})}{3+\lambda_{1,u}}&0\leq e[u]\leq 1,\\
1&e[u]>1,\\
0&e[u]<0,
\end{matrix}\right.\label{form23}
\end{align}
and $\lambda_{1,u}$ is expressed as
\begin{align}
&\lambda_{1,u}=\max\left\{0,-2(3+\tilde{e}[u]+\hat{e}[u]+\bar{e}[u]-\kappa(\lambda_{\tilde{e}[u]}+\lambda_{\hat{e}[u]}+\right.\nonumber\\
&\lambda_{\bar{e}[u]})),\left.2(\tilde{e}[u]+\hat{e}[u]+\bar{e}[u]-\kappa(\lambda_{\bar{e}[u]}+\lambda_{\tilde{e}[u]}+\lambda_{\hat{e}[u]}))\right\}. \label{form24}
\end{align}
Similarly, using the Lagrange multipliers method, the optimal $\alpha_{u}$, $c$, $\hat{\alpha}_{u}$, $\tilde{\eta}$ and $\tilde{\boldsymbol{x}}_{u,n_{1}}$ can be expressed as
\begin{align}
&\alpha_{u}=(2\tilde{\alpha}_{u}+2\kappa\lambda_{\tilde{\alpha}_{u}}-\lambda_{2,u})/2,\nonumber \\&c=\lambda_{3}\sigma_{n}^{2}/2+\tilde{c}-\kappa\lambda_{\tilde{c}},\nonumber\\
&\hat{\alpha}_{u}=\lambda_{4,u}/2+\tilde{\alpha}_{u}\tilde{c}-\kappa\lambda_{\hat{\alpha}_{u}},\nonumber\\&\gamma_{u}=(\tilde{\gamma}_{u}+\boldsymbol{q}^{H}\boldsymbol{b}_{u}\beta_{u}-\kappa(\lambda_{\tilde{\gamma}_{u}}+\lambda_{\gamma_{u}}))/4, \nonumber\\
&\tilde{\eta}=\frac{\hat{\eta}+\kappa\lambda_{\hat{\eta}}+\eta(\bar{\eta}-\kappa\lambda_{\bar{\eta}})+\bar{\eta}(\bar{\eta}+\kappa\lambda_{\bar{\eta}})}{1+\eta^{2}+\bar{\eta}^{2}}, \nonumber\\
&\tilde{\boldsymbol{x}}_{u,n_{1}}=(\boldsymbol{x}_{u,n_{1}}-\boldsymbol{x}_{u,n_{2}})-\boldsymbol{\lambda}_{\tilde{\boldsymbol{x}}_{u,n_{1}}}+\lambda_{5,k}\tilde{\boldsymbol{x}}_{u,n_{1}}^{(\tilde{t}-1)},\nonumber\\
&\tilde{\boldsymbol{y}}_{u,n_{1}}=(\boldsymbol{y}_{u,n_{1}}-\boldsymbol{y}_{u,n_{2}})-\boldsymbol{\lambda}_{\tilde{\boldsymbol{y}}_{u,n_{1}}}+\lambda_{6,k}\tilde{\boldsymbol{y}}_{u,n_{1}}^{(\tilde{t}-1)}. \label{form25}
\end{align}

The Lagrange multipliers 
$\alpha_{u}$, $c$ and $\hat{\alpha}_{u}$ corresponding to the optimization problems for 
$\lambda_{2,u}$, $\lambda_{3}$, 
$\lambda_{4,u}$, $\lambda_{5,u}$, and $\lambda_{6,u}$ are introduced as follows:
\begin{align}
&\lambda_{2,u}=\max\{0,2\tilde{\alpha}_{u}+2\kappa\lambda_{\tilde{\alpha}_{u}}-2|\tilde{\gamma}_{u}|^{2}\},\nonumber\\&\lambda_{3}=\max\{0,\frac{2}{\sigma_{n}^{2}}(\frac{P_{a}c}{\sigma_{n}^{2}}+\nonumber\kappa\lambda_{c}-c)\},\\&\lambda_{4,u}=\max\left\{0,2(\bar{e}[u]S_{u}^{2}+\kappa\lambda_{\hat{\alpha}_{u}}-\tilde{\alpha}_{u}c)\right\},\nonumber\\
&\lambda_{5,u}=\max\left\{0, (\|\tilde{\boldsymbol{x}}_{u,n_{1}}^{(\tilde{t}-1)}\|_{2}-v-2(\tilde{\boldsymbol{x}}_{u,n_{1}}^{(\tilde{t}-1)})^{T}((\boldsymbol{x}_{u,n_{1}}-\boldsymbol{x}_{u,n_{2}})\right.\nonumber\\
&\left.-\boldsymbol{\lambda}_{\tilde{\boldsymbol{x}}_{u,n_{1}}}))/(2\tilde{\boldsymbol{x}}_{u,n_{1}}^{(\tilde{t}-1)})^{T}\tilde{\boldsymbol{x}}_{u,n_{1}}^{(\tilde{t}-1)}\right\},\nonumber\\
&\lambda_{6,u}=\max\left\{0, (\|\tilde{\boldsymbol{y}}_{u,n_{1}}^{(\tilde{t}-1)}\|_{2}-v-2(\tilde{\boldsymbol{y}}_{u,n_{1}}^{(\tilde{t}-1)})^{T}((\boldsymbol{y}_{u,n_{1}}-\boldsymbol{y}_{u,n_{2}})\right.\nonumber\\
&\left.-\boldsymbol{\lambda}_{\tilde{\boldsymbol{y}}_{u,n_{1}}}))/(2\tilde{\boldsymbol{y}}_{u,n_{1}}^{(\tilde{t}-1)})^{T}\tilde{\boldsymbol{y}}_{u,n_{1}}^{(\tilde{t}-1)}\right\}.\label{form26}
\end{align}

The subproblem with respect to $\tilde{\boldsymbol{q}}$ can be written as
\begin{subequations}
\begin{align} 
\min_{\tilde{\boldsymbol{q}}}&|\boldsymbol{q}-\tilde{\boldsymbol{q}}+\kappa\boldsymbol{\lambda}_{\tilde{\boldsymbol{q}}}\|^{2},~\label{form27a}\\
\text{s.t.}~ 
&\|\tilde{\boldsymbol{q}}\|^{2}=1,~\label{form27b}
\end{align}\label{form27}
\end{subequations}
where subproblems (\ref{form27a}) and (\ref{form27b}) can be solved with the aid of projection method~\cite{cai2019robust}, with the solution given by
$\tilde{\boldsymbol{q}}=(\boldsymbol{q}+\kappa\boldsymbol{\lambda}_{\tilde{\boldsymbol{q}}})/\|\boldsymbol{q}+\kappa\boldsymbol{\lambda}_{\tilde{\boldsymbol{q}}}\|$.

The subproblem for $\boldsymbol{b}_{u}$ is given by
\begin{subequations}
\begin{align} 
\min_{\boldsymbol{b}_{u}}&\|\boldsymbol{a}(\boldsymbol{x}_{u},\boldsymbol{y}_{u})-\boldsymbol{b}_{u}+\kappa\boldsymbol{\lambda}_{\boldsymbol{b}_{u}}\|^{2}+\|\gamma_{u}-\boldsymbol{q}^{H}\boldsymbol{b}_{u}\beta_{u}+\kappa\lambda_{\gamma_{u}}\|^{2},\label{form28a}\\
\text{s.t.}~ 
&|\boldsymbol{b}_{u}(i)|=1.&\label{form28b}
\end{align}\label{form28}%    
\end{subequations}
Due to the separability of the unit modulus constraints, the one-iteration block coordinate descent (BCD)-type algorithm described in Appendix A of \cite{liu2021reconfigurable} can be effectively utilized to solve the problem (\ref{form28}).

\subsection{2-nd round of optimization}
For the subproblem over $\tilde{e}[u]$, $\hat{e}[u]$ and $\bar{e}[u]$, we continue to adopt the Lagrange multiplier method, and they are given by
\begin{align}
&\tilde{e}[u]=e[u]+\kappa\lambda_{\tilde{e}_{u}}-\lambda_{7}S_{u}/2, \nonumber\\
&\hat{e}[u]=e[u]+\kappa\lambda_{\hat{e}_{u}}-\lambda_{8}S_{u}/2, \nonumber\\
&\bar{e}[u]=e[u]+\kappa\lambda_{\bar{e}_{u}}-\lambda_{9,u}S_{u}^{2}/2.\label{form29}
\end{align}
The Lagrange multipliers $\lambda_{6}$, $\lambda_{7}$ and $\lambda_{8,u}$ are expressed as
\begin{align}
&\lambda_{7}=\max\{0,2[\sum\nolimits_{u=1}^{U}(e[u]+\kappa\lambda_{\tilde{e}_{u}})S_{u}-\hat{\eta}]/S_{u}^{2}\},\nonumber\\
&\lambda_{8}=\max\{0,2[\sum\nolimits_{u=1}^{U}(e[u]+\kappa\lambda_{\tilde{e}_{u}})S_{u}-\tilde{\eta}]/S_{u}^{2}\},\nonumber\\
&\lambda_{9,u}=\max\{0,2[(e[u]+\kappa\lambda_{\bar{e}_{u}}-S_{u}^{2}/\hat{\alpha}_{u})/S_{u}^{2}\}.\label{form30}
\end{align}
Then, using the gradient method, the optimal $\tilde{\alpha}_{u}$ can be expressed as
\begin{align}
&\tilde{\alpha}_{u}=(\alpha_{u}+\hat{\alpha}_{u}+\kappa(\lambda_{\tilde{\alpha}_{u}}+\lambda_{\hat{\alpha}_{u}}))/(1+\tilde{c}^{2}),\nonumber\\ 
&\hat{\eta}=(8M/K^{2}-2\tilde{\eta}-2\kappa\lambda\tilde{\eta})/(8/K^{2}-2),\nonumber\\
&\eta=c\sigma_{n}^{2}/(P_{a}\tilde{c}), ~\text{and}~\nonumber\\
&\boldsymbol{q}=(\boldsymbol{I}+\sum_{u=1}^{U}(\boldsymbol{q}^{H}\boldsymbol{b}_{u}\beta_{u})^{H}\boldsymbol{q}^{H}\boldsymbol{b}_{u}\beta_{u})^{-1}(\tilde{\boldsymbol{q}}-\kappa\boldsymbol{\lambda}_{\boldsymbol{q}}+(\gamma_{u}+\kappa\lambda_{\gamma_{u}})\nonumber\\
&\boldsymbol{q}^{H}\boldsymbol{b}_{u}\beta_{u}).\label{form31}
\end{align}

The subproblem with respect to $\boldsymbol{x}_{k}$ is given by
\begin{align}
\min_{\boldsymbol{x}_{u},\boldsymbol{y}_{u}}\|\angle\boldsymbol{a}(\boldsymbol{x}_{u},\boldsymbol{y}_{u})-\angle\boldsymbol{b}_{k}+\kappa\lambda_{\boldsymbol{b}_{k}}\|^{2}.\label{form32}      
\end{align}
Since $\boldsymbol{b}_{k}$ satisfies constant modulus-constrained, problem (\ref{form32}) is rewritten as
\begin{align}
&\mathcal{J}(\boldsymbol{x}_{u,n_{1}},\boldsymbol{y}_{u,n_{1}})=\sum\nolimits_{n_{1}'=1}^{N_{u}}(|\frac{2\pi}{\lambda}(\boldsymbol{\pi})^{T}\boldsymbol{\zeta}_{u,n_{1}'}-\angle\boldsymbol{b}_{u}(n_{1}')|^{2}\nonumber\\
&+\sum\nolimits_{n_{2}=1}^{N_{u}}\|\tilde{\boldsymbol{x}}_{u,n_{1}}-(\boldsymbol{x}_{u,n_{1}}-\boldsymbol{x}_{u,n_{2}})+\kappa\boldsymbol{\lambda}_{\tilde{\boldsymbol{x}}_{u,n_{1}}}\|^{2}\nonumber\\
&+\sum\nolimits_{n_{2}=1}^{N_{u}}\|\tilde{\boldsymbol{y}}_{u,n_{1}}-(\boldsymbol{y}_{u,n_{1}}-\boldsymbol{y}_{u,n_{2}})+\kappa\boldsymbol{\lambda}_{\tilde{\boldsymbol{y}}_{u,n_{1}}}\|^{2}).\label{form33}
\end{align}
Based on the first-order optimal condition, $\boldsymbol{x}_{u,n_{1}}$ is given by
\begin{align}
&(\boldsymbol{I}+\frac{4\pi^{2}}{\lambda^{2}}\boldsymbol{\pi}(\boldsymbol{\pi})^{T})\boldsymbol{x}_{u,n_{1}}-(\tilde{\boldsymbol{x}}_{u,n_{1}}+\sum_{n_{2}=1}^{N_{u}}\boldsymbol{x}_{u,n_{2}}+\nonumber\\
&\kappa\boldsymbol{\lambda}_{\tilde{\boldsymbol{x}}_{u,n_{1}}})-\frac{2\pi}{\lambda}\boldsymbol{\pi}\angle\boldsymbol{b}_{u}(n_{1})=0,\nonumber\\
&(\boldsymbol{I}+\frac{4\pi^{2}}{\lambda^{2}}\boldsymbol{\pi}(\boldsymbol{\pi})^{T})\boldsymbol{y}_{u,n_{1}}-(\tilde{\boldsymbol{y}}_{u,n_{1}}+\sum_{n_{2}=1}^{N_{u}}\boldsymbol{y}_{u,n_{2}}+\nonumber\\
&\kappa\boldsymbol{\lambda}_{\tilde{\boldsymbol{y}}_{u,n_{1}}})-\frac{2\pi}{\lambda}\boldsymbol{\pi}\angle\boldsymbol{b}_{u}(n_{1})=0.\label{form34}%
\end{align}
The solution of the problem in (\ref{form34}) is written as (\ref{form35}) at the top of this page.
\begin{figure*}

\begin{align}
&\boldsymbol{x}_{u,n_{1}}=\left\{\begin{matrix}
(\boldsymbol{I}+\frac{4\pi^{2}}{\lambda^{2}}\boldsymbol{\pi}(\boldsymbol{\pi})^{T})^{-1}((\tilde{\boldsymbol{x}}_{u,n_{1}}+\sum_{n_{2}=1}^{N_{u}}\boldsymbol{x}_{u,n_{2}}+\kappa\boldsymbol{\lambda}_{\tilde{\boldsymbol{x}}_{u,n_{1}}})+\frac{2\pi}{\lambda}\angle\boldsymbol{b}_{u}),&~\textrm{if}~ \boldsymbol{x}_{u,n_{1}}\in\mathcal{C}_{x}\\
\mathcal{C}_{x}^{max}, &~\textrm{else}~\textrm{if}~J(\mathcal{C}_{x}^{max})<J(\mathcal{C}_{x}^{min})\\
\mathcal{C}_{x}^{min}, &~\textrm{otherwise.}\\
\end{matrix}\right.\nonumber\\
&\boldsymbol{y}_{u,n_{1}}=\left\{\begin{matrix}
(\boldsymbol{I}+\frac{4\pi^{2}}{\lambda^{2}}\boldsymbol{\pi}(\boldsymbol{\pi})^{T})^{-1}((\tilde{\boldsymbol{y}}_{u,n_{1}}+\sum_{n_{2}=1}^{N_{u}}\boldsymbol{y}_{u,n_{2}}+\kappa\boldsymbol{\lambda}_{\tilde{\boldsymbol{y}}_{u,n_{1}}})+\frac{2\pi}{\lambda}\angle\boldsymbol{b}_{u}),&~\textrm{if}~ \boldsymbol{y}_{u,n_{1}}\in\mathcal{C}_{y}\\
\mathcal{C}_{y}^{max}, &~\textrm{else}~\textrm{if}~J(\mathcal{C}_{y}^{max})<J(\mathcal{C}_{y}^{min})\\
\mathcal{C}_{y}^{min}, &~\textrm{otherwise.}\\
\end{matrix}\right.,\label{form35}
\end{align}
\hrulefill
\end{figure*}

\subsection{3-rd round of optimization}
For the subproblem over $\bar{\eta}$ and $\tilde{c}$, we continue to adopt the Lagrange multiplier method, and they are given by
\begin{align}
&\bar{\eta}=\frac{\tilde{\eta}\tilde{c}-\eta\tilde{\eta}+\kappa(\tilde{\eta}\lambda_{\tilde{c}}+\lambda_{\bar{\eta}})}{1+\tilde{\eta}^{2}},\nonumber\\~\text{and}~
&\tilde{c}=\frac{c+\kappa\lambda_{\tilde{c}}-\tilde{\alpha}_{u}\hat{\alpha}_{u}-\kappa\lambda_{\hat{\alpha}_{u}}+\bar{\eta}\tilde{\eta}-\kappa\lambda_{\tilde{\eta}}}{1-\tilde{\alpha}_{u}^{2}}.\label{form36}
\end{align}

\subsection{Complexity Analysis}
To analyze the computational complexity of the proposed algorithm, we primarily focus on the updates of 
$\{e[u]\}$, 
$\{\alpha_{u},c,\hat{\alpha}_{u},\gamma_{u},\tilde{\eta},\tilde{\boldsymbol{x}}_{u,n_{1}},\tilde{\boldsymbol{y}}_{u,n_{1}}\}$,
$\{\tilde{\boldsymbol{q}}\}$,
$\{\boldsymbol{b}_{u}\}$,
\{$\tilde{e}[u]$,$\hat{e}[u]$,$\bar{e}[u]$\},
$\{\tilde{\alpha}_{u},\hat{\eta},\eta,\boldsymbol{q}\}$,
$\{\boldsymbol{x}_{u,n_{1}},\boldsymbol{y}_{u,n_{1}}\}$,
\{$\bar{\eta}$,$\tilde{c}$\}, which are the main determinants of complexity. 
The complexity of updating 
\{$e[u]$\} is $\mathcal{O}(U)$. When updating the $\{\boldsymbol{b}_{u}\}$, the complexity of one iteration of the proposed BCD-type algorithm is determined by $\mathcal{O}(U^{2})$. The complexity of updating 
\{$\tilde{u}_{k,k^{\prime}}$,$\tilde{\nu}_{k}$,$\boldsymbol{u}_{k}$\} depends on the bisection method used to search for the Lagrange multiplier. 
The complexity of updating 
$\{\alpha_{u},c,\hat{\alpha}_{u},\gamma_{u},\tilde{\eta},\tilde{\boldsymbol{x}}_{u,n_{1}},\tilde{\boldsymbol{y}}_{u,n_{1}}\}$ is $\mathcal{O}(UN_{T}+3U+2)$. 
The complexity of updating 
\{$\tilde{e}[u]$,$\hat{e}[u]$,$\bar{e}[u]$\} is $\mathcal{O}(3U)$. The complexity of updating 
$\{\tilde{\alpha}_{u},\hat{\eta},\eta,\boldsymbol{q}\}$ is $\mathcal{O}(N_{T}+U+2)$. The complexity of updating 
$\{\boldsymbol{x}_{u,n_{1}},\boldsymbol{y}_{u,n_{1}}\}$ and 
\{$\bar{\eta}$,$\tilde{c}$\} is $\mathcal{O}(2N_{T}U+2)$. Consequently, the overall complexity of the proposed algorithm can be represented as $\mathcal{O}(T_{1}T_{2}(U(2N_{T}+5)+U+N_{T}+U^{2}+6))$, in which the maximum number of iterations for
the inner and outer loops are denoted by $T_{1}$ and $T_{2}$.

\section{Bounding the Maximum Channel Gain and Convergence Implications}
\label{subsec:channel_bound}

In OTA FL, communication noise significantly impacts the accuracy of model aggregation, since the received signal at the server is a superposition of user gradients distorted by fading channels and additive noise. Let $\boldsymbol{e}_{2,t}$ denote the aggregation error at iteration $t$, defined as the difference between the ideal summed gradient and what the server receives. The mean squared error (MSE) of the aggregated gradient is given by
\begin{equation}
\mathbb{E}\bigl[\|\boldsymbol{e}_{2,t}\|_2^2\bigr]
~=~
\frac{K \,\sigma_n^2 \,\|\boldsymbol{g}_u\|^2}%
     {P_a \,\max\limits_{u \in \mathcal{U}_{t}^{s}}
      \bigl|\boldsymbol{q}_t^H \,\boldsymbol{h}_{u,t}(\boldsymbol{x}_{u,t},\boldsymbol{y}_{u,t})\bigr|^2},
\label{eq:mse_main}
\end{equation}
where \(K = |\mathcal{U}_{t}^{s}|\) is the number of participating users, $P_a$ is the transmit power constraint, and $\boldsymbol{q}_t$ is the beamforming vector applied at the server. To minimize this aggregation error, the beamformer $\boldsymbol{q}_t$ and user antenna positions $(\boldsymbol{x}_{u,t},\boldsymbol{y}_{u,t})$ can be jointly optimized, along with an appropriate selection of transmitting users $\mathcal{U}_{t}^{s}$.

\paragraph{Maximum Channel Gain Bound}

By leveraging the Cauchy-Schwarz inequality and the field response matrix (FRM), we can show that
\begin{equation}
  \label{eq:channel_bound}
  \max_{u \in \mathcal{U}_{t}^{s}}
  \bigl|\boldsymbol{q}_t^H \,\beta_{u,t}\,\boldsymbol{a}_{u,t}(\boldsymbol{x}_{u,t},\boldsymbol{y}_{u,t})\bigr|^2
  \;\le\;
  \max_{u \in \mathcal{U}_{t}^{s}}\|\beta_{u,t}\|^2\,N_T,
\end{equation}
where $\beta_{u,t}$ encapsulates path-loss/attenuation factors, and $\boldsymbol{a}_{u,t}(\boldsymbol{x}_{u,t},\boldsymbol{y}_{u,t})$ is the array steering vector dependent on user antenna positions. The term $N_T$ denotes the number of antennas in use. This result implies that $\boldsymbol{q}_t$ is subjected to $\|\boldsymbol{q}_t\|^2=1$, then the effective channel gain cannot exceed the product $\|\beta_{u,t}\|^2\, N_T$. 

An important insight is that one can approach this upper bound by carefully aligning the beamformer phases $\theta_{q_{t, i}}$ with the user antenna positions $(\boldsymbol{x}_{u,t},\boldsymbol{y}_{u,t})$. For instance, under the far-field model, if
\begin{align}
\left|\frac{\theta_{q_{t,j}}-\theta_{q_{t,i}}}{2 \cos(\theta_{u,t})}\right| > v_{x},~\text{and}~
\left|\frac{\theta_{q_{t,j}}-\theta_{q_{t,i}}}{2 \sin(\phi_{u,t})}\right| > v_{y},
\end{align}
for given thresholds $v_x$ and $v_y$, the array response terms add constructively, allowing the effective channel gain $\bigl|\boldsymbol{q}_t^H \boldsymbol{h}_{u,t}\bigr|$ to approach the theoretical limit in \eqref{eq:channel_bound}. Hence, while $\|\beta_{u,t}\|^2\, N_T$ is a hard constraint, combining precise antenna positioning with phase alignment can push the actual channel magnitude closer to this bound. The details of the proof are in Appendix~\ref{appendix-1}.

\paragraph{Convergence Implications}
We analyze how communication noise, modeled by $\boldsymbol{e}_{2,t}$ in \eqref{eq:mse_main}, affects the convergence of federated learning under $L$-smooth and $\mu$-strongly convex loss functions. In particular, it is shown that the expected suboptimality of the global model $F(\boldsymbol{w}_T)$ at iteration $T$ can be bounded by
\begin{equation}
\mathbb{E}[F(\boldsymbol{w}_T) - F(\boldsymbol{w}^*)]
\;\leq\;
\phi_T \,\mathbb{E}[F(\boldsymbol{w}_0) - F(\boldsymbol{w}^*)]
\;+\;
\Psi_T,
\label{eq:convergence_main}
\end{equation}
where both $\phi_T$ and $\Psi_T$ depend on the additional variance introduced by $\boldsymbol{e}_{2,t}$. As the maximum channel gain increases (and thus $\mathbb{E}[\|\boldsymbol{e}_{2,t}\|^2]$ decreases), one obtains faster convergence rates and lower final training loss. Consequently, the joint optimization of beamforming vectors, user positions, and user selection is instrumental in improving transmission performance and enhancing convergence performance in OTA FL. The details of the proof are in Appendix~\ref{appendix-2}.

\section{Numerical Results} 
\label{experiments}

In this section, we extensively evaluate the proposed approach in a multi-user OTA federated learning scenario, under two commonly used image-classification datasets: \emph{MNIST} dataset~\cite{deng2012mnist} and \emph{CIFAR-10} dataset~\cite{Krizhevsky09learningmultiple}. Throughout all experiments, we consider $U=100$ users and $N_{u}$ antennas. The distance $d_{u}$ between user $u$ and the parameter server is uniformly distributed as $d_{u}\sim U[10\,\text{m},100\,\text{m}]$, and the path loss follow the COST Hata model $\mathrm{PL}[dB]=139.1+35.22\log\bigl(d_{u}[\text{km}]\bigr)$. $\mathrm{PL}$ is a distance-dependent factor that scales the variance of the small-scale fading, ensuring the channel gain remains zero-mean Gaussian but with power inversely proportional to the path loss. Each user’s channel is
$  \boldsymbol{h}_{u,t} = \boldsymbol{h}_{u} 
  \sim 
  \mathcal{CN}\bigl(\boldsymbol{0}, \tfrac{1}{\mathrm{PL}}\boldsymbol{I}\bigr),
$
which is assumed constant (i.e., static channels) over the training process. The maximum average transmit power is set to $P_{a}=0\,\text{dBm}$, and the receiver noise power is $\sigma_{n}^{2}=-20\,\text{dBm}$, accounting for thermal noise and interference. 

\paragraph{Training and Model Setup}
We use a global learning rate of $\lambda=0.05$ for parameter updates. Each device employs its entire local data batch to compute gradients in every communication round (i.e., one local epoch per global iteration). Consequently, the number of rounds equals the number of global epochs.

\paragraph{Benchmark Methods}
We compare our PDD and FA approaches with both the traditional maximum ratio transmission (MRT) scheme and the fixed point algorithm (FPA)-based approach:
\begin{itemize}
    \item \textbf{Select All}: All $U$ devices transmit in every round. Beamforming and antenna positioning follow the approach of \cite{xu2023accelerating}, which optimizes these parameters to improve communication efficiency in FL. However, this method does not jointly optimize device selection, which limits its adaptability in dynamic network environments.
    \item \textbf{DC Method}: Beamforming and device selection are jointly optimized via difference-of-convex (DC) programming to keep the global model’s MSE below a threshold $J$ \cite{xu2023accelerating}. While this approach effectively reduces MSE, it does not incorporate dynamic antenna positioning, which can further enhance communication performance.
\end{itemize}
% Additionally, we consider the MRT (Maximum Ratio Transmission) scheme as a baseline for beamforming design, and the RFA (Random Fluid Antenna) plus APS (Alternating Position Selection) methods~\cite{hao2024fluid} for comparison with other fluid-antenna strategies.

Additionally, MRT serves as a fundamental beamforming approach by aligning each transmit antenna’s signal with the channel’s complex conjugate, thereby maximizing the received power at the intended user~\cite{zhou2024fluid}. Meanwhile, random fluid antenna (RFA) leverages spatial flexibility by reconfiguring the antenna within a limited region to exploit small-scale fading differences and improve signal strength, and the alternating position selection (APS) method~\cite{hao2024fluid} further refines this concept by quantizing possible antenna locations and iterating among them to find an advantageous position without the complexity of continuous optimization.

\subsection{MNIST Experiments}

\paragraph{Dataset and Device Configuration}
Each device holds $S_{u}=270$ images from the MNIST dataset. As in all experiments, each device calculates the gradient from its local batch per communication round. 

\begin{figure}[!ht]
    \centering
    \includegraphics[scale=0.45]{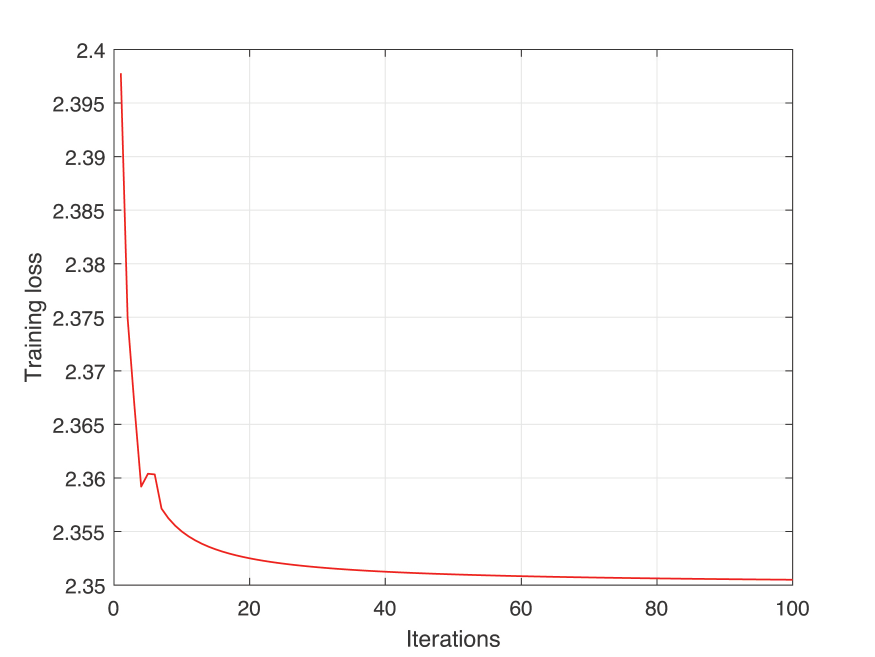}
    \captionsetup{justification=centering}
    \caption{Training loss over iterations (MNIST).}
    \label{FIGUREM5}
\end{figure}

\paragraph{Training Loss Convergence}
Fig.~\ref{FIGUREM5} shows the variation of training loss as the number of iterations grows. The proposed PDD algorithm achieves a smooth and consistent decline in training loss, converging roughly between the 10th and 20th iterations. This rapid stabilization indicates that the approach effectively handles noise and gradient aggregation. 

\begin{figure}[!ht]
    \centering
    \includegraphics[scale=0.45]{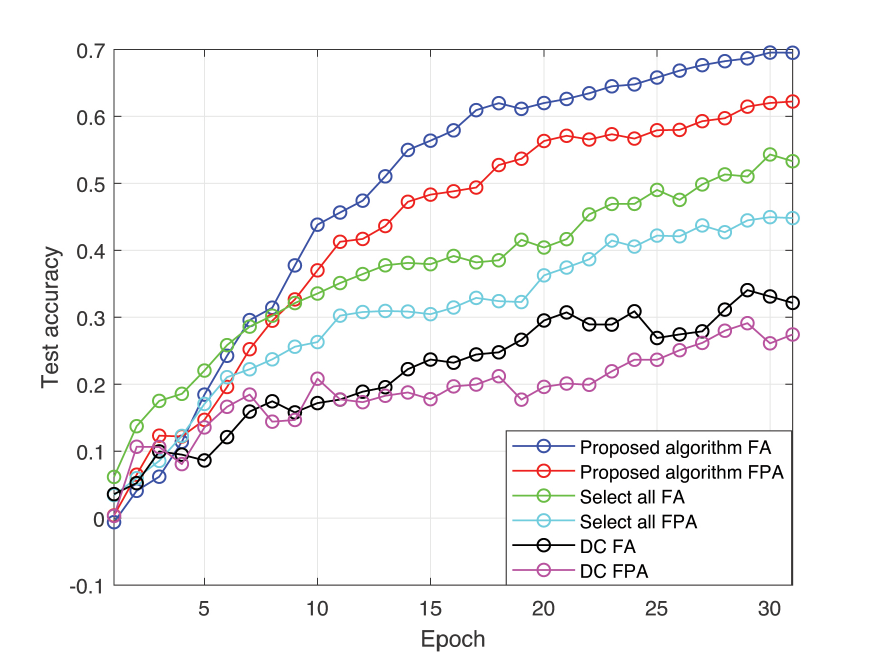}
    \captionsetup{justification=centering}
    \caption{Test accuracy over epochs (MNIST).}
    \label{FIGUREICC3}
\end{figure}

\begin{figure}[!ht]
    \centering
    \includegraphics[width=0.45\textwidth]{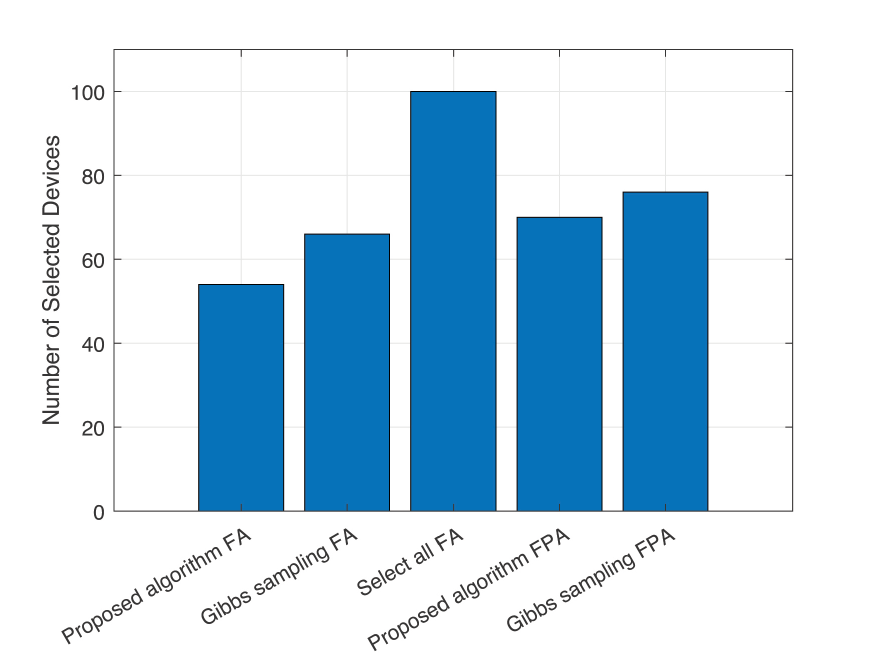}
    \captionsetup{justification=centering}
    \caption{Number of devices selected by each method (MNIST).}
    \label{FIGUREICC2}
\end{figure}

\paragraph{Test Accuracy and Device Selection}
Fig.~\ref{FIGUREICC3} depicts the test accuracy over communication rounds, averaged across 16 channel realizations. The FA-based scheme within our PDD framework converges to over 70\% test accuracy by around $T=25$, surpassing both “Select All” and the DC method by significant margins (around 40\% improvement over DC because antennas' positions are random for the DC method). Meanwhile, Fig.~\ref{FIGUREICC2} compares the number of selected devices across methods. Our method selects fewer devices than the FPA-based approaches, confirming that it balances user selection to mitigate noise while ensuring sufficient training data diversity. FA outperforms FPA because repositionable antennas achieve better channel conditions for the chosen users, thereby reducing noise and speeding up convergence.

\begin{figure}[!ht]
    \centering
    \includegraphics[scale=0.45]{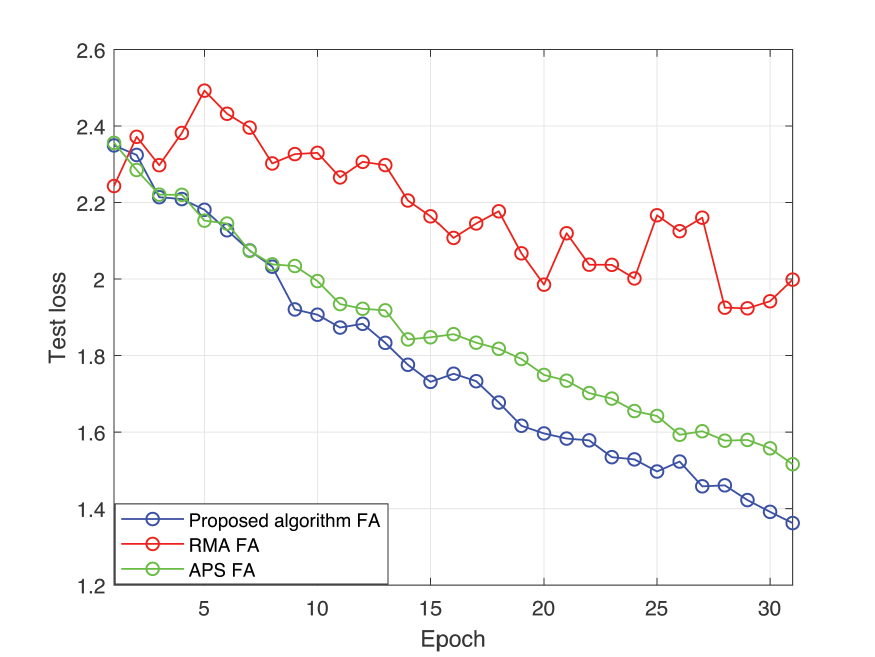}
    \captionsetup{justification=centering}
    \caption{Test loss over epochs (MNIST).}
    \label{FIGUREICC4}
\end{figure}

\paragraph{Test Loss Comparison}
As shown in Fig.~\ref{FIGUREICC4}, our PDD-based FA algorithm consistently achieves the lowest test loss among the considered approaches. In contrast, RFA positioning and the APS method exhibit higher and more fluctuating test loss. Their reliance on random or quantized antenna positioning does not reliably account for multi-user interference and channel conditions, thus degrading their performance. 

\paragraph{Beamforming Schemes Comparison}

\begin{figure}[!ht]
    \centering
    \includegraphics[scale=0.45]{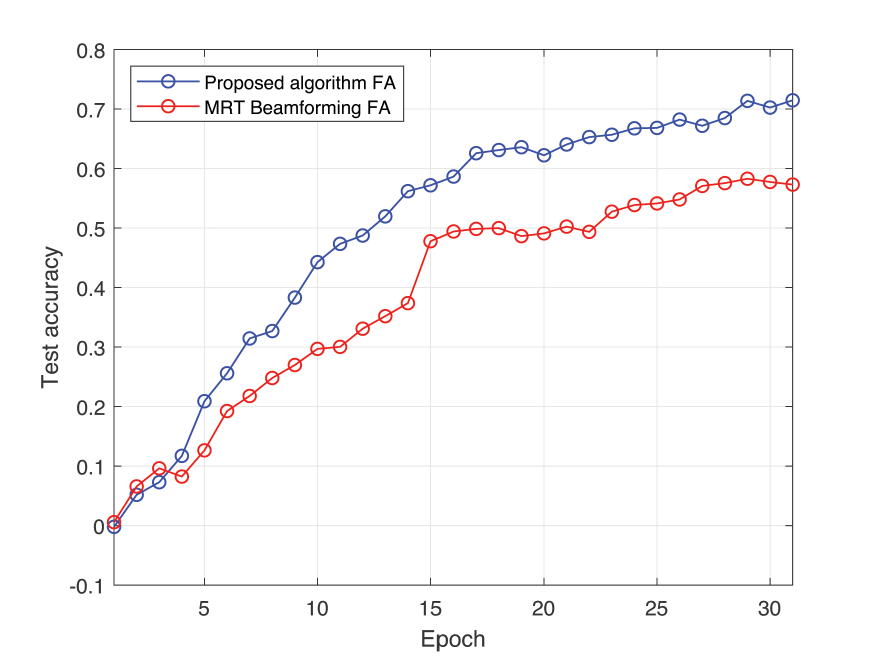}
    \captionsetup{justification=centering}
    \caption{Proposed beamforming design versus MRT (MNIST).}
    \label{FIGUREICC5}
\end{figure}

In Fig.~\ref{FIGUREICC5}, we compare the performance of the proposed beamforming design with the traditional MRT scheme. The results demonstrate that the proposed scheme significantly outperforms MRT in terms of testing accuracy. This performance gap arises because MRT optimizes only the signal direction by maximizing the received signal power without considering interference from other users. While this approach is effective in single-user scenarios, it becomes suboptimal in multi-user uplink communication settings, where interference management is crucial. Since our study focuses on a multi-user environment, MRT exhibits relatively poor performance compared to the proposed scheme, which is explicitly designed to mitigate interference and optimize overall system performance.

\subsection{CIFAR-10 Experiments}

\paragraph{Dataset and Device Configuration}
We also evaluate the same OTA FL framework on the more challenging CIFAR-10 dataset. Each user $u$ again holds 270 i.i.d.\ samples across various classes, and the same transmit power/noise level setup is used. 

\begin{figure}[!ht]
    \centering
    \includegraphics[scale=0.45]{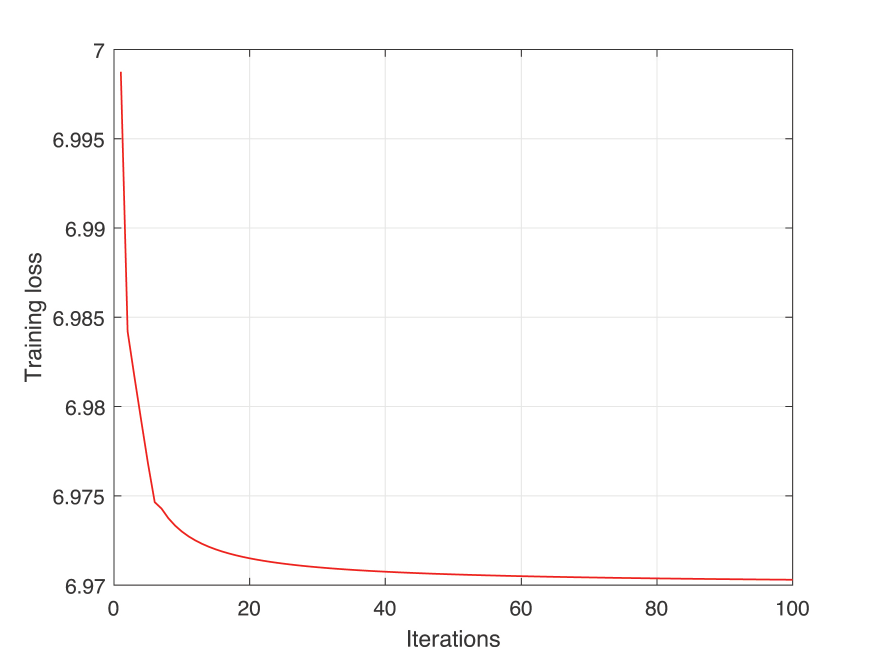}
    \captionsetup{justification=centering}
    \caption{Training loss over iterations (CIFAR-10).}
    \label{FIGUREC5}
\end{figure}

\paragraph{Training Loss Convergence}
Fig.~\ref{FIGUREC5} shows the training loss versus iteration. The PDD algorithm exhibits a consistent decline, converging between the 10th and 20th iterations despite more complex image features. 

\begin{figure}[!ht]
    \centering
    \includegraphics[scale=0.45]{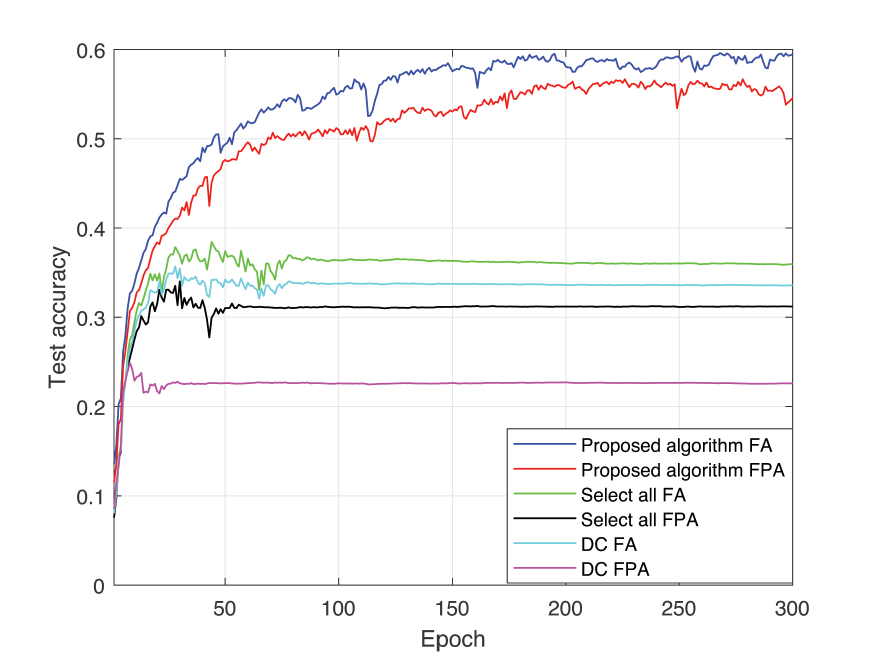}
    \captionsetup{justification=centering}
    \caption{Test accuracy over epochs (CIFAR-10).}
    \label{JSAC__1}
\end{figure}

\paragraph{Test Accuracy}
The average test accuracy for various methods is shown in Fig.~\ref{JSAC__1}. All FA-based schemes (including ours) outperform the FPA-based counterparts, attaining up to a 50\% gain over the “Select All” and DC baselines. These latter two methods converge but exhibit lower final accuracy, likely due to higher aggregated noise (Select All) or insufficiently flexible selection constraints. 

\begin{figure}[!ht]
    \centering
    \includegraphics[scale=0.45]{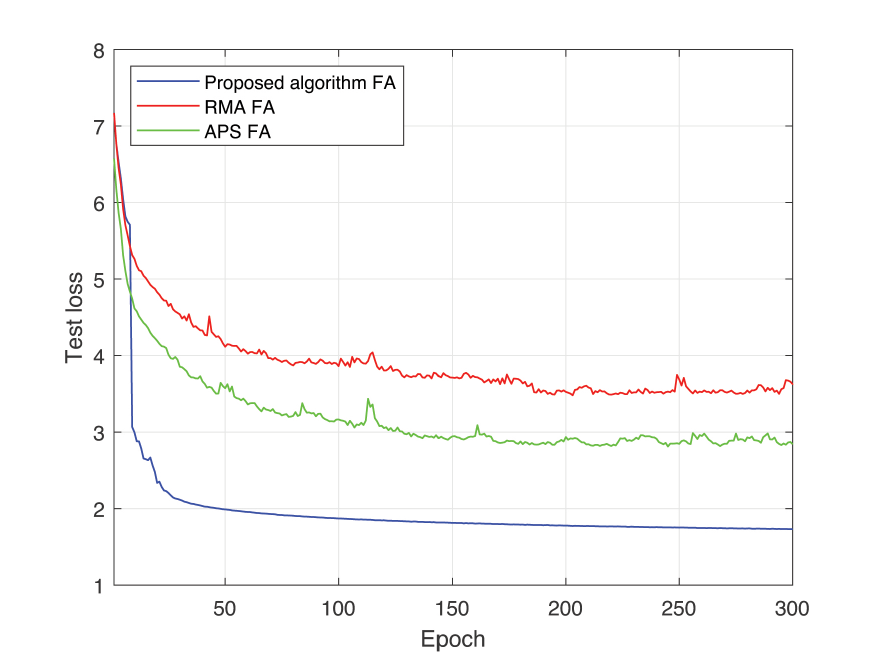}
    \captionsetup{justification=centering}
    \caption{Test loss over epochs (CIFAR-10).}
    \label{JSAC__2}
\end{figure}

\paragraph{Test Loss}
Fig.~\ref{JSAC__2} compares the test loss under different beamforming and antenna-positioning approaches. Our proposed PDD remains the best performer, maintaining low and stable loss throughout training. By contrast, the RFA and APS methods have higher and more fluctuating test loss.

\begin{figure}[!ht]
    \centering
    \includegraphics[width=0.45\textwidth]{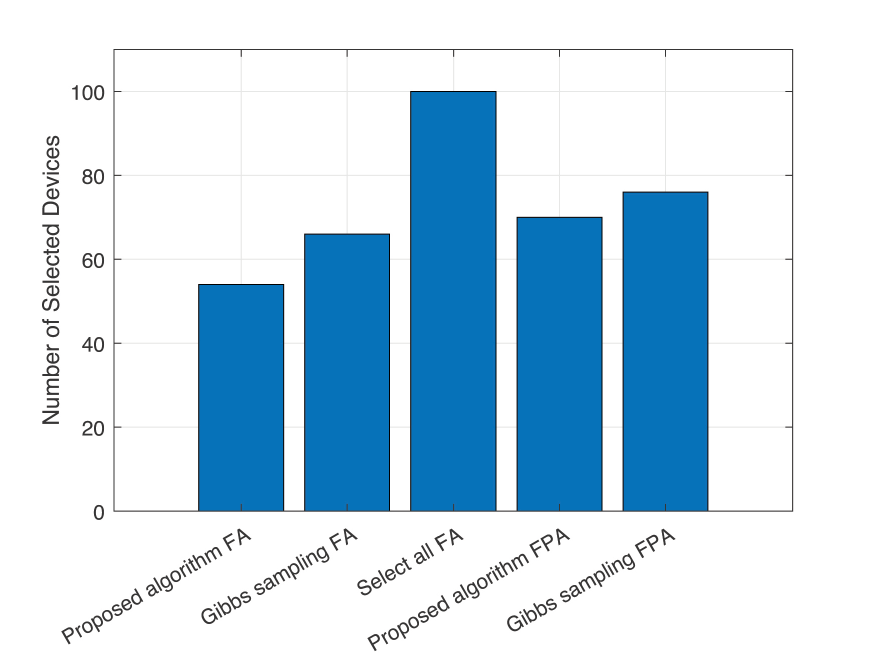}
    \captionsetup{justification=centering}
    \caption{Number of selected devices versus FA methods (CIFAR-10).}
    \label{JSAC_4}
\end{figure}

\paragraph{Device Selection}
Fig.~\ref{JSAC_4} presents the average number of devices selected and its standard deviation, measured across 20 channel realizations. Similar to MNIST, the FA scheme in our method selects fewer devices than FPA methods, indicating a more refined balance between noise resilience and training data availability.

\paragraph{Beamforming Scheme Comparison}
Finally, Fig.~\ref{JSAC_3} compares our beamforming approach with MRT. The proposed algorithm again achieves significantly higher test accuracy, reinforcing that MRT’s single-user focus fails in the presence of multi-user interference, whereas our solution is designed to optimize aggregated signals in such a multi-user FL context. Across both MNIST and CIFAR-10 datasets, our PDD method stands out in terms of:

\begin{figure}[!ht]
    \centering
    \includegraphics[scale=0.45]{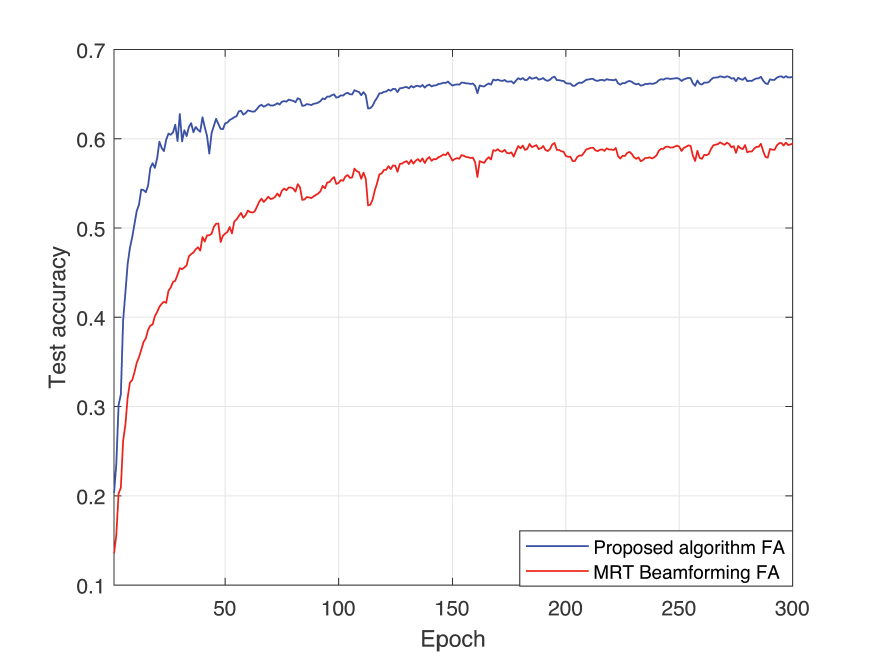}
    \captionsetup{justification=centering}
    \caption{Proposed beamforming versus MRT for CIFAR-10.}
    \label{JSAC_3}
\end{figure}

\begin{itemize}
\item Efficient Convergence: Rapid decrease in training loss within 10--20 iterations, attributed to effective beamforming and antenna-positioning coordination.
\item Higher Test Accuracy: Consistently achieves superior classification performance compared to DC, Select All, RFA, APS, and MRT baselines.
\item Adaptive Device Selection: Selects fewer devices yet mitigates OTA noise effectively, balancing SNR and dataset coverage.
\item Robust to Channel Fluctuations: Maintains stable performance across diverse channel realizations and noise conditions.
\end{itemize}
Overall, these results confirm that jointly optimizing beamforming, antenna positioning, and device selection yield substantial performance gains in both training speed and inference accuracy for OTA FL.

\section{Conclusions} \label{conclusion}

In this paper, we collaboratively optimize uplink receiver beamforming, FA positioning, and device selection in FA-enabled OTA FL to minimize global training loss after any arbitrary $T$ communication rounds, accounting for time-varying wireless channels. To solve this non-convex problem, we use a training loss upper bound based on existing references and devise strategies for receiver beamforming, FA positioning, and device selection to minimize this bound. Our approach utilizes PDD and SCA. The PDD framework applies an alternating optimization method to address the subproblems of device selection and receiver beamforming iteratively, and SCA is employed to handle the non-convex FA-positioning constraints. We also present an efficient algorithm that solves the device selection subproblem with minimal complexity.

\bibliographystyle{ieeetr}
\bibliography{related}

\begin{appendices}
\section{Communication Error and Joint Optimization}\label{appendix-1}

Let \(K = |\mathcal{U}_{t}^{s}|\) be the number of selected users, each with a local gradient \(\boldsymbol{g}_{u} \in \mathbb{C}^{D}\) assumed to have the same or a worst-case norm, and each able to transmit with power \(P_{a}\), i.e., \(|a_{u,t}|^2 \le P_{a}\). The server, equipped with \(M\) antennas, receives \(\boldsymbol{r}_{t} = \sum_{u \in \mathcal{U}_{t}^{s}} \boldsymbol{h}_{u,t}(a_{u,t}\,\boldsymbol{g}_{u}) + \boldsymbol{n}_{t}\), where \(\boldsymbol{n}_{t}\sim\mathcal{CN}(\mathbf{0}, \sigma_{n}^2 \mathbf{I}_{M})\) and each channel \(\boldsymbol{h}_{u,t}(\boldsymbol{x}_{u,t}, \boldsymbol{y}_{u,t})\) depends on FA positions. A unit-norm beamformer \(\boldsymbol{q}_{t}\) is applied, and the server scales the output by \(\eta_{t} = 1/\max_{u}\lvert \boldsymbol{q}_{t}^H \boldsymbol{h}_{u,t}\rvert\) to produce \(\hat{\boldsymbol{g}}_{t} = \eta_{t} \,\boldsymbol{q}_{t}^H \boldsymbol{r}_{t}\). Defining the aggregation error \(\boldsymbol{e}_{2,t} = \sum_{u \in \mathcal{U}_{t}^{s}}\boldsymbol{g}_{u} - \hat{\boldsymbol{g}}_{t}\), one finds that under full power \(|a_{u,t}|^2 = P_{a}\) and uniform gradient norms.

Hence, we have the MSE of OTA gradient aggregation
\begin{align}
    \mathbb{E}\bigl[\|\boldsymbol{e}_{2,t}\|_2^2\bigr]
~=~
\frac{K \,\sigma_n^2 \,\|\boldsymbol{g}_u\|^2}%
     {P_a \,\max\limits_{u \in \mathcal{U}_{t}^{s}}
      \bigl|\boldsymbol{q}_t^H \,\boldsymbol{h}_{u,t}(\boldsymbol{x}_{u,t},\boldsymbol{y}_{u,t})\bigr|^2}.\label{formA1}
\end{align}

A higher \(\max_{u}|\boldsymbol{q}_t^H \boldsymbol{h}_{u,t}(\cdot)|^2\) reduces the error denominator, thereby minimizing \(\mathbb{E}[\|\boldsymbol{e}_{2,t}\|_2^2]\). According to the channel model in (\ref{form4}), $\max\limits_{u \in \mathcal{U}_{t}^{s}}\bigl|\boldsymbol{q}_t^H \,\boldsymbol{h}_{u,t}(\boldsymbol{x}_{u,t},\boldsymbol{y}_{u,t})\bigr|^2$ is rewritten as
\begin{align}
\max\limits_{u \in \mathcal{U}_{t}^{s}}\bigl|\boldsymbol{q}_t^H \,\beta_{u,t}\boldsymbol{a}_{u,t}(\boldsymbol{x}_{u,t},\boldsymbol{y}_{u,t})\bigr|^2.\label{formA2}
\end{align}
Using the definition of the beamformer and array elements, $\bigl|\boldsymbol{q}_t^H \,\beta_{u,t}\boldsymbol{a}_{u,t}(\boldsymbol{x}_{u,t},\boldsymbol{y}_{u,t})\bigr|^2$ is expanded as
\begin{align}
&|\beta_{u,t}|^{2}\bigl|\sum_{i=1}^{N_{T}}q_{t,i}e^{j\frac{2\pi}{\lambda}(x_{u,t,i}\cos(\theta_{u,t})+y_{u,t,i}\sin(\phi_{u,t}))}\bigr|^{2}=|\beta_{u,t}|^{2}\times\nonumber\\
&(\sum_{i=1}^{N_{T}}|q_{t,i}|^{2}+\sum_{i\neq j}^{N_{T}}2\mathrm{Re}\{q_{t,i}q_{t,j}e^{j\frac{2\pi}{\lambda}(x_{u,t,i}\cos(\theta_{u,t})+y_{u,t,i}\sin(\phi_{u,t}))}\nonumber\\
&e^{j\frac{2\pi}{\lambda}(x_{u,t,j}\cos(\theta_{u,t})+y_{u,t,j}\sin(\phi_{u,t}))}\})=|\beta_{u,t}|^{2}(1+\sum_{i\neq j}^{N_{T}}2\mathrm{Re}\{q_{t,i}q_{t,j}\nonumber\\
&e^{j\frac{2\pi}{\lambda}(x_{u,t,i}\cos(\theta_{u,t})+y_{u,t,i}\sin(\phi_{u,t}))}\nonumber\\
&e^{j\frac{2\pi}{\lambda}(-x_{u,t,j}\cos(\theta_{u,t})-y_{u,t,j}\sin(\phi_{u,t}))}\}).\label{formA3}
\end{align}
For maximizing (\ref{formA3}), one might want
\begin{align}
&\theta_{q_{t,i}}+\frac{2\pi}{\lambda}(x_{u,t,i}\cos(\theta_{u,t})+y_{u,t,i}\sin(\phi_{u,t}))=\theta_{q_{t,j}}\nonumber\\
&+\frac{2\pi}{\lambda}(x_{u,t,j}\cos(\theta_{u,t})+y_{u,t,j}\sin(\phi_{u,t})).\label{formA4}
\end{align}
The equation in (\ref{formA4}) is transformed as
\begin{align}
&(\theta_{q_{t,i}}-\theta_{q_{t,j}})+\frac{2\pi}{\lambda}((x_{u,t,i}-x_{u,t,j})\cos(\theta_{u,t})\nonumber\\&+(y_{u,t,i}-y_{u,t,j})\sin(\phi_{u,t}))=0.\label{formA5}
\end{align}
To maximize the expression in (\ref{formA3}), let $x_{u,t,i}-x_{u,t,j}=0$ and $y_{u,t,i}-y_{u,t,j}=0$, the solution of equation in (\ref{formA5}) is given by
\begin{align}
x_{u,t,i}-x_{u,t,j}=\frac{(\theta_{q_{t,j}}-\theta_{q_{t,i}})}{2\cos(\theta_{u,t})},\nonumber\\
y_{u,t,i}-y_{u,t,j}=\frac{(\theta_{q_{t,j}}-\theta_{q_{t,i}})}{2\sin(\phi_{u,t})}.
\end{align}
Therefore, based on the above discussion, we can derive the following conclusion
when $|\frac{(\theta_{q_{t,j}}-\theta_{q_{t,i}})}{2\cos(\theta_{u,t})}|>v_{x}$ and $|\frac{(\theta_{q_{t,j}}-\theta_{q_{t,i}})}{2\sin(\phi_{u,t})}|>v_{y}$, we can obtain the lower bound of $
\mathbb{E}\bigl[\|\boldsymbol{e}_{2,t}\|_2^2\bigr]$.

\section{Gradient Descent Convergence with Communication Noise}~\label{appendix-2}

If $F(\boldsymbol{w})$ is $L$-smooth, then the following inequality holds:
\begin{align}
&F(\boldsymbol{w}_{t+1})
\leq
F(\boldsymbol{w}_t)
+ \nabla F(\boldsymbol{w}_t)^\top
        \bigl(\boldsymbol{w}_{t+1}-\boldsymbol{w}_t\bigr)
+ \frac{L}{2}\,\nonumber\\
&\times\|\boldsymbol{w}_{t+1}-\boldsymbol{w}_t\|^2.    
\end{align}
Substituting the update rule $\boldsymbol{w}_{t+1} = \boldsymbol{w}_t - \eta g_t$, we obtain:
\begin{align}
F(\boldsymbol{w}_{t+1})
\leq
F(\boldsymbol{w}_t)
- \eta\,\nabla F(\boldsymbol{w}_t)^\top g_t
+ \frac{L\,\eta^2}{2}\,\|g_t\|^2.
\end{align}
Taking expectation conditioned on $\boldsymbol{w}_t$, we get:
\begin{align}
\mathbb{E}[F(\boldsymbol{w}_{t+1})]
\leq
\mathbb{E}[F(\boldsymbol{w}_t)]
- \eta\,\mathbb{E}[\nabla F(\boldsymbol{w}_t)^\top g_t]
+ \frac{L\,\eta^2}{2}\,\mathbb{E}[\|g_t\|^2].
\end{align}
If the local gradient estimator is unbiased, i.e.,
\begin{align}
\mathbb{E}[g_t \mid \boldsymbol{w}_t] = \nabla F(\boldsymbol{w}_t),
\end{align}
then it follows that
\begin{align}
\mathbb{E}[\nabla F(\boldsymbol{w}_t)^\top g_t] = \|\nabla F(\boldsymbol{w}_t)\|^2.
\end{align}
Using variance decomposition, we expand
\begin{align}
\mathbb{E}[\|g_t\|^2]
=
\|\nabla F(\boldsymbol{w}_t)\|^2
+ \mathbb{E}[\|g_t - \nabla F(\boldsymbol{w}_t)\|^2].
\end{align}
Substituting into (2) gives
\begin{align}
&\mathbb{E}[F(\boldsymbol{w}_{t+1})]
\leq
\mathbb{E}[F(\boldsymbol{w}_t)]
- \eta\,\|\nabla F(\boldsymbol{w}_t)\|^2\nonumber\\
&+ \frac{L\,\eta^2}{2}
\Bigl(
  \|\nabla F(\boldsymbol{w}_t)\|^2
  + \sigma^2
  + \mathbb{E}[\|\boldsymbol{e}_{2,t}\|^2]
\Bigr),
\end{align}
where $\sigma^2$ is the variance of the stochastic gradient noise, and $\|\boldsymbol{e}_{2,t}\|$ represents the additional error due to communication.

If $F(\boldsymbol{w})$ is $\mu$-strongly convex, we use the inequality:
\begin{align}
\|\nabla F(\boldsymbol{w}_t)\|^2
\geq
2\mu
\bigl(
  F(\boldsymbol{w}_t)
  - F(\boldsymbol{w}^*)
\bigr).    
\end{align}
Substituting (3) into the previous bound yields:
\begin{align}
&\mathbb{E}[F(\boldsymbol{w}_{t+1})]
\leq
\mathbb{E}[F(\boldsymbol{w}_t)]
- 2\mu \eta
\bigl(
  \mathbb{E}[F(\boldsymbol{w}_t)] - F(\boldsymbol{w}^*)
\bigr)\nonumber\\
&+ \frac{L\,\eta^2}{2}
\bigl(
  \sigma^2 + \mathbb{E}[\|\boldsymbol{e}_{2,t}\|^2]
\bigr).
\end{align}
Rearranging, we obtain
\begin{align}
\mathbb{E}[F(\boldsymbol{w}_{t+1}) - F(\boldsymbol{w}^*)]
\leq
\phi_t \mathbb{E}[F(\boldsymbol{w}_t) - F(\boldsymbol{w}^*)]
+ \psi_t,
\end{align}
where
\begin{align}
\phi_t
=
1 - \frac{\mu}{L} \bigl(1 - \alpha\,r(\boldsymbol{q}_t, \boldsymbol{e}_t)\bigr),
\quad
\psi_t
=
\frac{L\,\eta^2}{2} \bigl(\sigma^2 + r(\boldsymbol{q}_t, \boldsymbol{e}_t)\bigr).    
\end{align}
Here, $r(\boldsymbol{q}_t, \boldsymbol{e}_t)$ represents the penalty due to communication noise, modeled as
\begin{align}
r(\boldsymbol{q}_t, \boldsymbol{e}_t)
=
\frac{K\,\sigma_n^2}{P_a \max\limits_{u\in \mathcal{U}_{t}^{s}} |\boldsymbol{q}_t^H\boldsymbol{h}_{u,t}(\boldsymbol{x}_{u,t},\boldsymbol{y}_{u,t})|^2}.    
\end{align}
Expanding over $T$ iterations, we obtain
\begin{align}
&\mathbb{E}[F(\boldsymbol{w}_T) - F(\boldsymbol{w}^*)]
\leq
\Biggl(\prod_{t=0}^{T-1} \phi_t\Biggr)
\mathbb{E}[F(\boldsymbol{w}_0) - F(\boldsymbol{w}^*)]\nonumber\\
&+ \sum_{t=0}^{T-1}
  \Biggl(\prod_{\tau=t+1}^{T-1} \phi_\tau\Biggr)
  \psi_t.
\end{align}
This result shows that the convergence rate depends directly on the communication error $\boldsymbol{e}_{2,t}$, which is affected by the beamformer $\boldsymbol{q}_t$ and the channel conditions. By jointly optimizing $\boldsymbol{q}_t$ and antenna positions $x_{u,t}$ to maximize $\max_{u\in \mathcal{U}_{t}^{s}}\lvert\boldsymbol{q}_t^H\boldsymbol{h}_u(x_{u,t})\rvert$, thereby reducing noise in OTA aggregation and accelerates convergence. Iteratively refining both $\boldsymbol{q}_t$ and $x_{u,t}$ over multiple rounds strengthens the effective channel gain, highlighting how well-designed beamforming and antenna configurations improve the overall convergence rate in communication-efficient FL.

\end{appendices}

\end{document}